\documentclass[sigplan,nonacm]{acmart}

\usepackage{xspace}
\usepackage[ruled,linesnumbered,vlined]{algorithm2e}
\usepackage{booktabs}
\usepackage{tabularx}
\usepackage{xcolor}
\usepackage{placeins}

\newcommand{\sys}{\textsc{CircLS}\xspace}

\begin{document}

\title{CircLS: Compiling Lattice Surgery to Physical Circuits with Dynamic Allocation}

\author{John Yuehan Zhang}
\authornote{Email: yuehanzhang6666@gmail.com}

\begin{abstract}

In fault-tolerant quantum computing, lattice surgery (LS) is one
of the leading ways to realize logical operations, and the Pauli
product measurement (PPM) is the basic instruction of LS-based
computing. Compilers that work on the PPM sequence, however,
stay at the logical level rather than the physical circuit level.
This is because lowering PPMs from the logical level to physical circuits is
complicated. \textbf{\sys} is a full-stack lattice
surgery compiler: it compiles a quantum program through its PPM
sequence to a stim circuit, lowering every PPM through linear-time
stabilizer construction rules. On this basis, we develop a
compiler that allocates data patches dynamically: each patch is
allocated at its first use and freed at its last use, and the
freed tiles can be reused as ancilla paths. Against the two baselines,
\sys reduces the allocated spacetime volume by 34\% and 27\%
and the logical error rate (LER) by 40\% and 48\%, and it compiles
programs that they cannot.  \sys is open source at
\url{https://github.com/John-YuehanZhang/CircLS}.
\end{abstract}

\maketitle

\section{Introduction}\label{sec:intro}

On quantum hardware, physical qubits are too noisy to run useful
algorithms directly~\cite{preskill2018nisq,gidneyekera2021factor}.  Quantum error
correction (QEC) closes this gap by encoding each logical qubit
redundantly across many physical
qubits to suppress
errors, enabling fault-tolerant quantum computing
(FTQC)~\cite{dennis2002topological,fowler2012surface}.  Experiments have demonstrated quantum
memory~\cite{ryananderson2021realtime,krinner2022realizing,google2023suppressing,acharya2025belowthreshold}, and attention has
turned to logical
operations~\cite{erhard2021entangling,bluvstein2024logical,
wang2026sclsprocessor}.  \emph{Lattice surgery} (LS) is one of the
leading ways to realize logical
operations~\cite{horsman2012lattice,fowler2018low,litinski2019game},
as demonstrated in recent
experiments~\cite{besedin2026lattice,wang2026sclsprocessor}.

The basic instruction of LS-based computing is the \emph{Pauli product
measurement} (PPM)~\cite{bravyi2016trading,litinski2019game}: a joint
measurement on a chosen set of logical qubits,
carried out by LS along an ancilla path connecting their
patches.  A practical program executes a vast number of PPMs: a
single quantum chemistry instance can require hundreds of billions
of PPMs~\cite{beverland2022req}.

The PPMs of a program differ widely from each other.  First, the
number of patches a PPM measures, called its \emph{weight}, is not
fixed.  Second, the bases can mix: each patch can be measured in the $X$,
$Y$, or $Z$ basis.
Third, the ancilla path is not unique: the same PPM can be routed
along different paths.  Finally, even with the ancilla path fixed,
different stabilizer constructions can realize the PPM.  As a
consequence, realizing arbitrary PPMs at the circuit level is
complex.

The complexity is compounded by a further requirement: a PPM must
be lowered to the physical circuit without breaking fault
tolerance.  Here, fault tolerance means constructing stabilizers
that implement the joint patch measurement while preserving the
code distance $d$.  Doing so is highly non-trivial.

Because of this complexity, the evaluation of PPM-based computing
remains at the logical level, without a realistic cost model from
physical circuits.  Some compilers place the patches and route the
ancilla
paths~\cite{lao2018mapping,beverland2022edpc,molavi2025dascot,
hamada2026routing}; others optimize the PPM sequence
itself~\cite{litinski2019game,kan2025sparo,o3ls2026,puremagic2025,
sethi2026ppr}; both stop without emitting a physical circuit.  The
metrics are logical-level quantities, and compiler optimizations
are therefore performed on these logical-level metrics.  Moreover,
the logical error rate (LER) is predicted by untested
formulas~\cite{o3ls2026,kan2025sparo,litinski2019game,
beverland2022req} rather than measured on physical circuits.  On
our benchmarks, the predictions of the composition
formula~\cite{o3ls2026} deviate from the measured LER by 26 to
52\% in the median.

\begin{figure*}
  \centering
  \includegraphics[width=\textwidth]{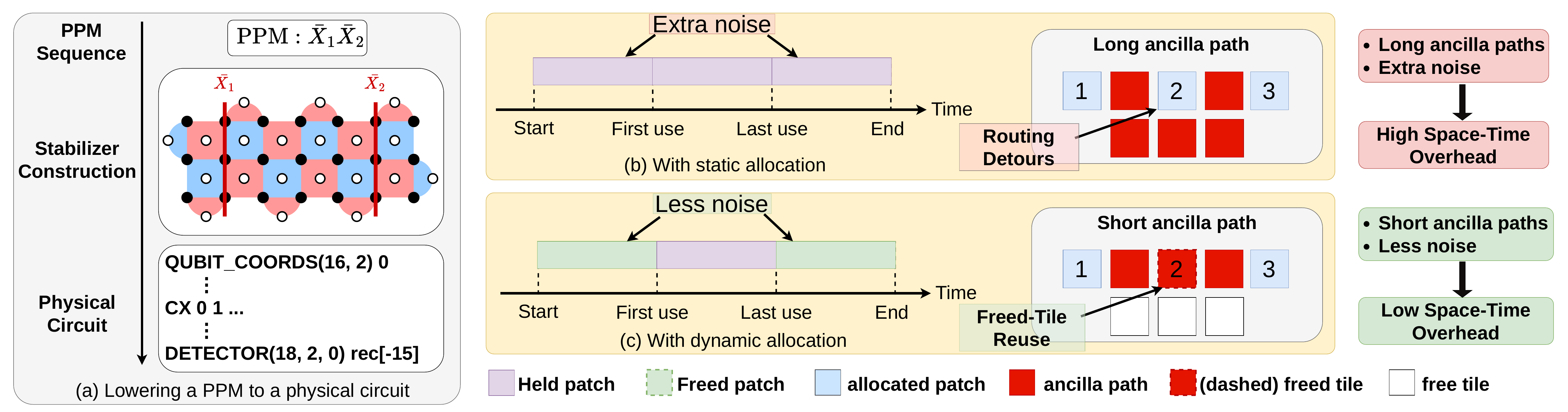}
  \caption{(a)~Lowering the PPM $\bar{X}_1\bar{X}_2$: \sys
    constructs the stabilizers that measure the product, then emits
    the physical circuit.  (b)~Static allocation: a data patch holds
    its tile from start to end, so it collects extra noise before its
    first use and after its last use, and ancilla paths must detour
    around held tiles; the path that measures patches 1 and 3 takes
    five tiles.  (c)~Dynamic allocation: \sys allocates the patch at
    its first use and frees it at its last use, so the patch collects
    less noise, and the freed tile is open for ancilla path reuse;
    the same path runs straight through three tiles.}
  \label{fig:overview}
\end{figure*}

Our key insight is that local correctness guarantees global
correctness: as long as every local stabilizer construction is
right, the global stabilizer construction is right.  Lowering
therefore reduces to a small set of local rules.

Based on this insight, we build \sys, a full-stack lattice surgery
compiler that takes a quantum program through its PPM sequence to
a stim circuit.  Its core is a unified, linear-time stabilizer
construction algorithm that lowers an arbitrary PPM sequence to a
stim circuit
(Section~\ref{sec:framework}).  The construction guarantees fault tolerance:
we verify that the code distance remains $d$ on every construction
shape.  This lowering provides a key piece of infrastructure for
the FTQC era: PPM-based
programs can now be run, verified, and measured as physical
circuits.  Based on this lowering, \sys can obtain the LER the way
an experiment would: it samples the compiled circuit under noise,
decodes the outcomes, and counts the failures, so every LER comes
from real samples rather than formulas.

Second, we develop a physical-level compiler using dynamic
allocation (Section~\ref{sec:lifetime}).  Existing compilers keep every data patch allocated
from start to
finish~\cite{hamada2026routing,molavi2025dascot,o3ls2026,
topols2026,bharadwaj2026qomet} (Figure~\ref{fig:overview}(b)).  This static
allocation has two drawbacks: the held tiles force ancilla paths
to detour, so the paths get longer; and a patch is exposed to
noise when unnecessary, before its first use and after its last
use.  With dynamic allocation, a data patch is allocated at its
first use and freed at its last use: the freed tiles can be reused as
ancilla paths, which become shorter, and the patch collects no
noise before allocation or after freeing
(Figure~\ref{fig:overview}(b) and (c)).  Beyond allocation, we
design a mapper that places the patches so that the ancilla paths
stay short, and a router that draws the ancilla path of every PPM
through the free tiles (Section~\ref{sec:mapping-routing}).

In summary, this paper makes three contributions:
\begin{itemize}
\item \textbf{Unified lowering algorithm.}  We design a unified,
  linear-time algorithm that lowers PPM sequences to stim circuits,
  letting compilers' output be verified at the circuit level and
  its LER measured from real samples.
\item \textbf{Dynamic patch allocation.}  A patch is allocated at
  its first use, freed at its last use, and its space can be reused
  by ancilla paths; we design a physical-level compiler around it.
\item \textbf{Comprehensive evaluation.}  Against two prior
  toolchains, \sys reduces the
  allocated spacetime volume by 34\% and 27\% and the LER by
  40\% and 48\%, reduces compile time by 98\% and 96\%, and compiles
  programs that they cannot.
\end{itemize}

\section{Background}\label{sec:background}

This section covers what the rest of the paper builds on: lattice
surgery, which measures Pauli products on surface-code patches
(Section~\ref{sec:ls-ppm}); the compilation pipeline from a
quantum circuit to a stim circuit (Section~\ref{sec:pipeline}); and
routing and stabilizer construction
(Section~\ref{sec:construction-bg}).

\subsection{Lattice Surgery and Pauli Product Measurements}\label{sec:ls-ppm}

A surface-code \emph{patch} is one logical qubit built from
$O(d^2)$ physical qubits, where $d$ is the code
distance~\cite{dennis2002topological,fowler2012surface}.  A
\emph{tile} is the space for one patch~\cite{litinski2019game}.

\begin{figure}
  \centering
  \includegraphics[width=\columnwidth]{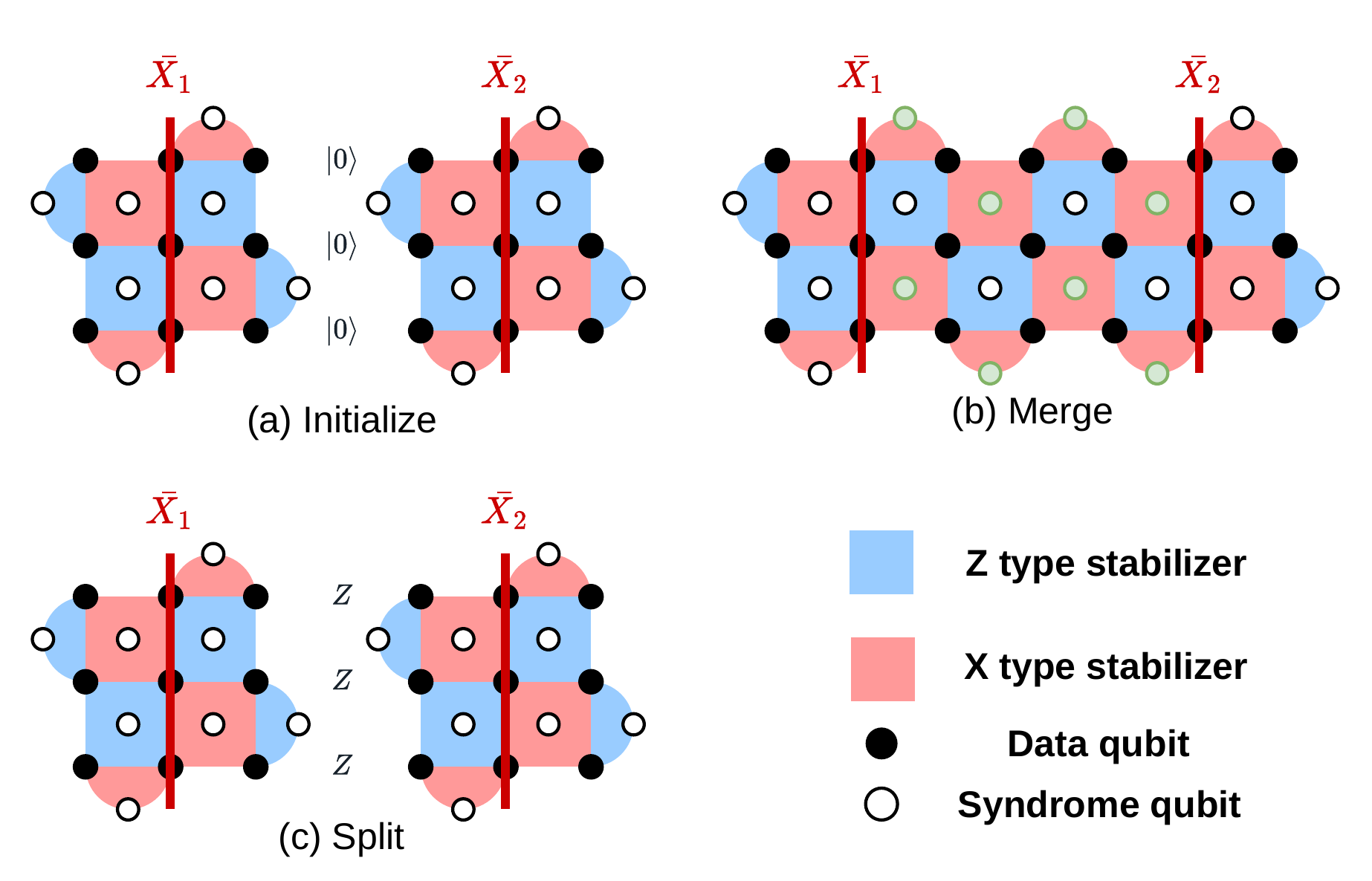}
  \caption{LS for measuring $\bar{X}_1\bar{X}_2$.
(a) Initialize the intermediate qubits between two patches. (b) Merge the patches: the product of the green stabilizers equals $\bar{X}_1\bar{X}_2$. (c) Split the patches by measuring the intermediate qubits in the
$Z$ basis.}
  \label{fig:ls-steps}
\end{figure}

LS implements logical operations by merging and
splitting
patches~\cite{horsman2012lattice,fowler2018low,litinski2019game}.  Its
basic use is a joint measurement of logical Pauli operators, such as
$\bar{X}_1\bar{X}_2$.  Figure~\ref{fig:ls-steps} shows the three
steps.  First, the intermediate qubits are initialized in $|0\rangle$.
Second, the two patches are merged into one patch by
measuring the stabilizers of the intermediate region. Third, the patches are split: the intermediate
qubits are measured in the $Z$ basis, and the outcomes update the
classically tracked Pauli frame.  The example above is simple.  However, practical computations
need mixed products, such as $\bar{X}_1\bar{Z}_2$, and joint
measurements over more than two patches, such as
$\bar{X}_1\bar{Z}_2\bar{X}_3$.  Realizing these general PPMs is
much more complex.

\begin{figure}
  \centering
  \includegraphics[width=\columnwidth]{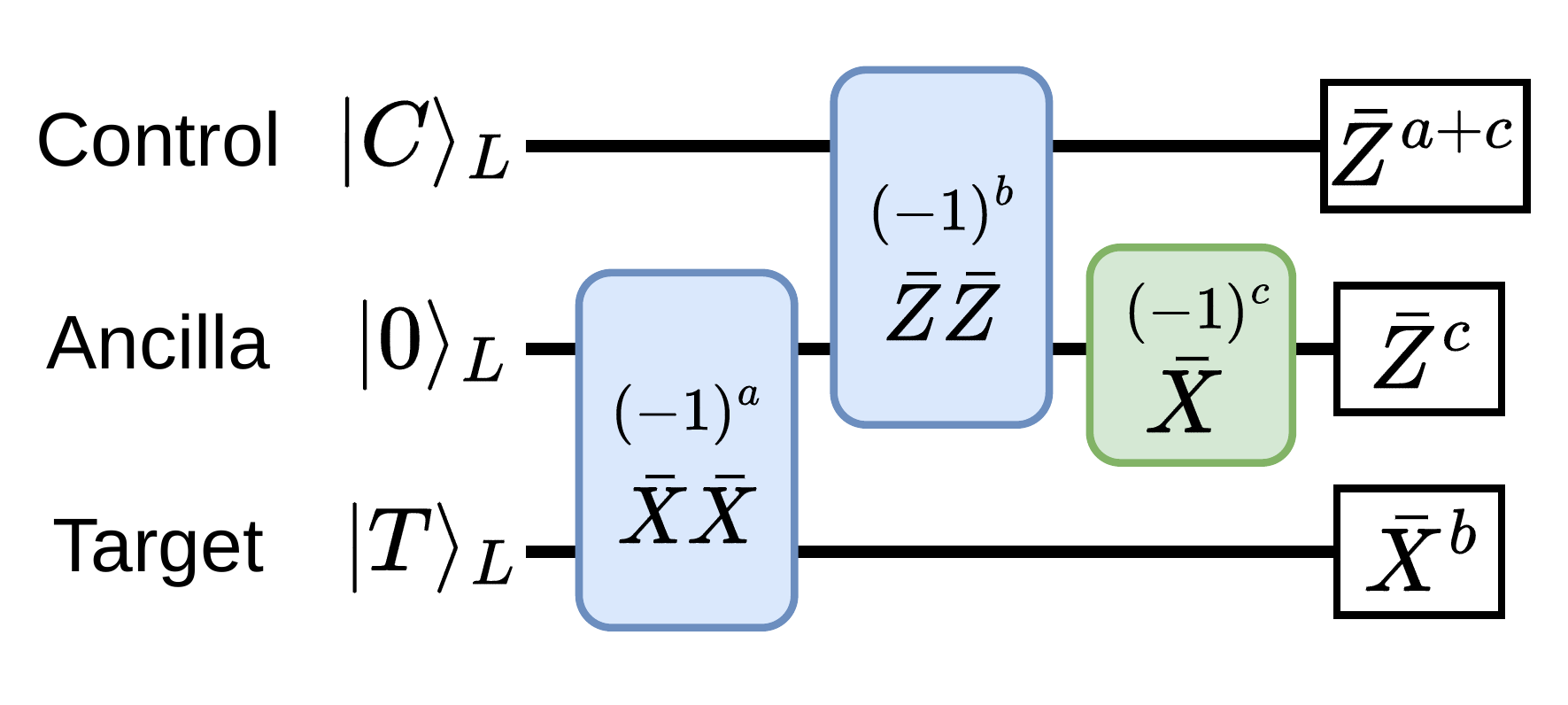}
  \caption{Logical CNOT by LS: three PPMs
    with measurement results $a$, $b$, $c$, followed by three Pauli
    corrections.}
  \label{fig:ls-cnot}
\end{figure}

Through the merge and split, LS performs a PPM on the
measured patches~\cite{horsman2012lattice,litinski2019game}.  PPMs can
implement various logical
operations~\cite{bravyi2016trading,litinski2019game,litinski2019magic,fowler2018low},
such as the CNOT: Figure~\ref{fig:ls-cnot} shows the three PPMs
and the Pauli corrections that implement it.  In fact, with
similar methods, and with magic states as inputs, PPMs can implement
universal quantum
computing~\cite{bravyi2016trading,litinski2019game,fowler2018low,litinski2019magic}.

\subsection{The Compilation Pipeline}\label{sec:pipeline}

The path from an algorithm to an executable circuit has
three layers.  At the top, the algorithm is a quantum circuit in QASM~\cite{cross2017openqasm}.  At the bottom, the executable form
is a stim circuit~\cite{gidney2021stim}.

The middle layer is an intermediate representation of the same
computation, and it comes in two patterns.  One pattern is the PPM
sequence: the circuit is rewritten as a list of PPMs.  The other
pattern is the ZX diagram: a graph intermediate
representation~\cite{coecke2011interacting,vandewetering2020zxcalculus,
kissinger2020pyzx}.
The ZX diagram is compiled in 3D: the computation is modeled as
connected 3D blocks, and the optimization works on this 3D
model~\cite{tqec2026joss,hao2025breakeven}. However, some compilers use the
PPM sequence as their intermediate
representation~\cite{litinski2019game,kan2025sparo,
o3ls2026,puremagic2025}, not the ZX diagram. Also, some operators of the
3D route are unimplemented, such as the $Y$-basis block, so some
quantum programs cannot be compiled. 

\sys provides full-stack compilation: it takes a PPM sequence,
maps and routes it, and lowers it to a stim circuit.

\subsection{Routing and Stabilizer Construction}\label{sec:construction-bg}

Realizing one PPM on the lattice takes two steps: routing and
stabilizer construction.

Routing finds an \emph{ancilla path}: a tree of free tiles that
connects the measured patches.  It has one constraint and one
goal: the ancilla path cannot occupy tiles held by data patches,
and the ancilla path should be as short as possible.  However,
existing methods use static
allocation~\cite{hamada2026routing,molavi2025dascot,o3ls2026,topols2026}:
every data patch holds its tile for the whole computation.
Ancilla paths must then detour around the held tiles, so they get
longer.  A longer ancilla path raises the spacetime volume and
the LER, and it blocks other ancilla paths.  This is why \sys
uses dynamic allocation.

Stabilizer construction builds stabilizers on the ancilla path,
turning the patches and the ancilla path into one merged patch
satisfying three requirements.  First, all stabilizers of the
merged patch must commute with each other.  Second, the merged patch
must have the right number of stabilizers.  A stabilizer code on $D$
data qubits with $S$ independent stabilizers encodes $D-S$ logical
qubits~\cite{gottesman1997stabilizer}.
Merging $n$ data patches leaves $n-1$ logical qubits, so the
merged patch must satisfy $D-S=n-1$.  Third, the merged patch must
measure the requested product: the product of some of the new
stabilizers must equal the measured operator.  The three
requirements make the merged patch a valid patch that measures the requested
operator.

However, satisfying the three requirements does not imply fault tolerance.  To
verify fault tolerance, the merged patch itself must be checked:
its code distance must remain $d$.

\section{Dynamic Patch Allocation}\label{sec:lifetime}

\begin{figure*}
  \centering
  \includegraphics[width=\textwidth]{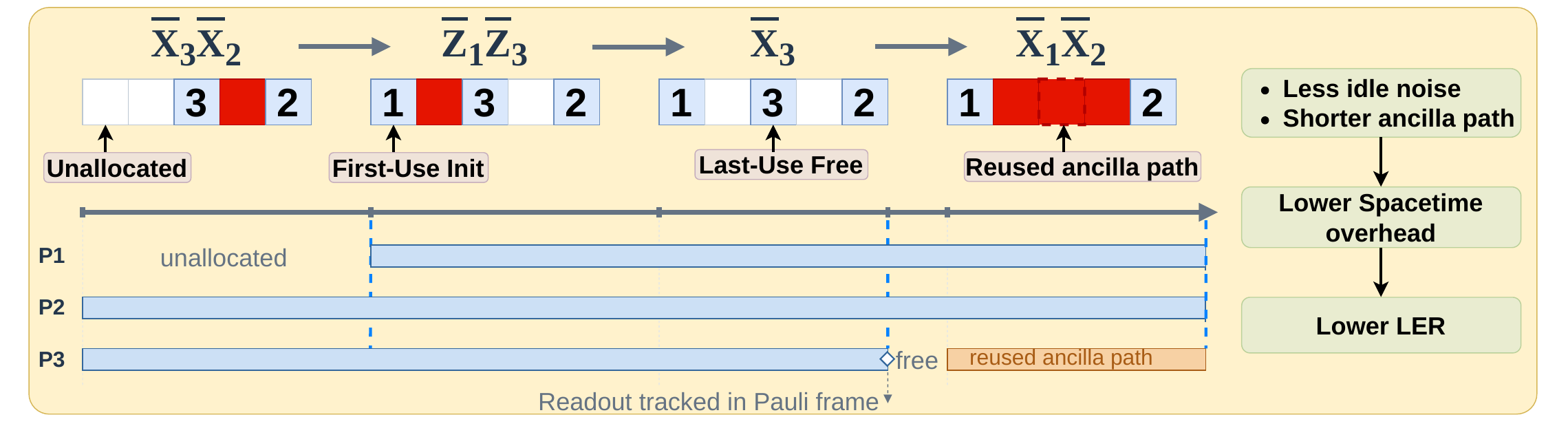}
  \caption{Dynamic allocation on one PPM sequence:
    $\mathrm{PPM}_1=\bar{X}_3\bar{X}_2$,
    $\mathrm{PPM}_2=\bar{Z}_1\bar{Z}_3$,
    $\mathrm{PPM}_3=\bar{X}_3$,
    $\mathrm{PPM}_4=\bar{X}_1\bar{X}_2$.}
  \label{fig:dynalloc}
\end{figure*}

\sys targets a compiled circuit with a smaller spacetime volume
and a lower LER.  The compiler pursues these goals in two ways:
it shrinks the space each PPM occupies, and it shortens the time each data patch stays
allocated.

Every stage of the pipeline in this section is classical and runs
at compile time.  The allocation follows two rules: a patch is
initialized at its first use (Section~\ref{sec:firstuse}) and freed
at its last use (Section~\ref{sec:lastuse}), so its live range runs
from first use to last use.  Section~\ref{sec:benefits} explains
what the shorter live range buys: the freed tiles can serve as
ancilla paths, and the patch collects less idle noise.
Figure~\ref{fig:dynalloc} shows an example of dynamic allocation
for a PPM sequence.

\subsection{First-Use Initialization}\label{sec:firstuse}

Existing compilers initialize every data patch at the start of the
program.  \sys initializes a data patch at the first PPM that uses
it.  As an example, take the PPM sequence of
Figure~\ref{fig:dynalloc}: $\bar{X}_3\bar{X}_2$,
$\bar{Z}_1\bar{Z}_3$, $\bar{X}_3$, $\bar{X}_1\bar{X}_2$.  Patch 1
first appears in the second PPM, so it is initialized there, not
at the start.  Before its
first use a patch holds no qubits, blocks no tile, and runs no
syndrome extraction.

An $|i\rangle$ ancilla patch is special: its fault-tolerant
initialization needs extra rounds~\cite{gidney2024inplacey}, while
$|0\rangle$ and $|+\rangle$ need only one round.  \sys starts the
initialization of an $|i\rangle$ patch early, overlapped with the
rounds of the preceding PPM, so the extra rounds add no time.

\subsection{Last-Use Freeing}\label{sec:lastuse}

Existing compilers measure every data patch at the end of the
program.  \sys frees a scratch patch right after its last PPM:
the patch is measured, and its tile is freed.  In
Figure~\ref{fig:dynalloc}, patch 3 last appears in the third PPM,
so it is freed there, not at the end.  The readout bit of a freed
patch is needed only for later Pauli corrections, which \sys
tracks classically in the Pauli frame, so freeing the patch loses
nothing.

An \emph{output patch} is different: it holds a logical qubit
whose final state is part of the program's output.  \sys keeps
its tile allocated until the end of the program.

Together the two rules define the live range of a patch as the
interval from its first use to its last use.  Outside that
interval, its tile is not occupied.

\subsection{Benefits of Dynamic Allocation}\label{sec:benefits}

A shorter live range offers two benefits.

\vspace{2pt}\noindent\textbf{\textit{Freed tiles can be reused as
ancilla paths.}}  The router treats the tile of an unallocated or
freed patch as an ordinary free tile.  A later PPM can run its
ancilla path through it, as the last PPM does in
Figure~\ref{fig:dynalloc}.  If every patch held its tile from the
start of the program to the end, the ancilla paths would have to
detour around the held tiles and would grow longer.

\vspace{2pt}\noindent\textbf{\textit{Less idle noise.}}  A patch
is exposed to noise throughout its live range, whether or not a
PPM touches it.  Under static allocation, every patch remains
allocated during the whole program, so this exposure includes the
time before its first use and after its last use.  Dynamic
allocation removes this exposure before first use.  For scratch
patches, it also removes exposure after last use.  This reduces
exposure to idle noise and lowers the LER.

\section{Mapping and Routing}\label{sec:mapping-routing}

\begin{figure*}[h]
  \centering
  \begin{minipage}[b]{0.290\textwidth}\centering
    \includegraphics[width=\linewidth]{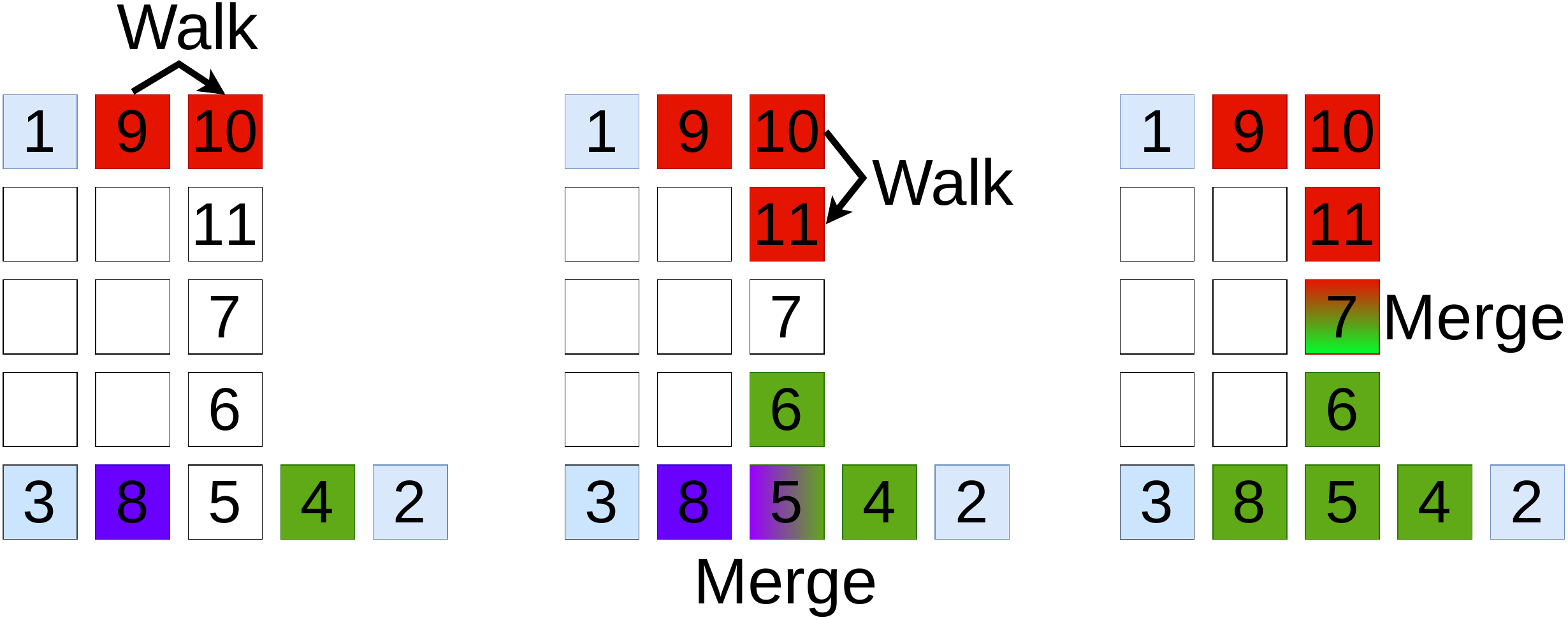}\\ (a)~Exact, for small $k$
  \end{minipage}\hfill
  \begin{minipage}[b]{0.690\textwidth}\centering
    \includegraphics[width=\linewidth]{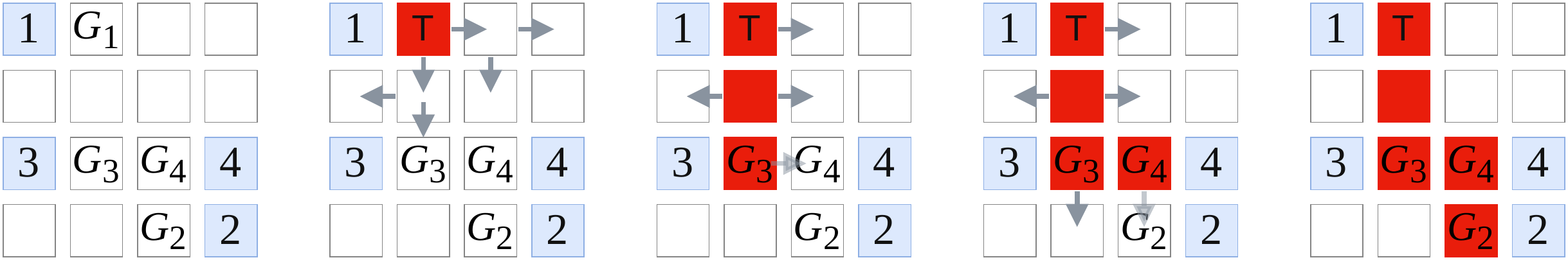}\\ (b)~Heuristic, for large $k$
  \end{minipage}
  \caption{The two cases of routing: (a)~the exact search for
    $k \le 9$, and (b)~the greedy heuristic used for $k > 9$
    (shown with $k=4$ for simplicity).}
  \label{fig:corridor-search}
\end{figure*}

With the PPM sequence and the live ranges fixed, two stages place
the sequence on the lattice.  Mapping decides where each patch sits
on the grid, aiming to make the ancilla paths across the sequence
as short as possible (Section~\ref{sec:mapping}).  Routing finds
an ancilla path for each PPM through the free tiles, including the
tiles of unallocated or freed patches.  For a PPM on a few patches, it uses an
exact search to find the shortest ancilla path.  For a larger PPM,
it uses a fast heuristic
(Section~\ref{sec:routing}).

\subsection{Mapping}\label{sec:mapping}

Mapping decides where each patch sits on the grid.  A patch can be
moved later, but moving is
expensive~\cite{litinski2019game,fowler2018low}, so \sys places
every patch once and does not move it during execution.  The placement must be good
from the start: the goal is to make the ancilla paths of the whole
sequence short.

Mapping works on a grid of tiles.  To leave room for ancilla paths,
\sys places data patches only on tiles whose row and column
numbers are both odd.  The remaining tiles are reserved for ancilla
paths, leaving free tiles on all four sides of each data patch.

A placement is scored by estimated ancilla path length.  For a PPM on
two patches $i$ and $j$ at tiles $(x_i, y_i)$ and $(x_j, y_j)$,
\sys estimates the ancilla path as
\begin{equation*}
L = |x_i - x_j| + |y_i - y_j| - 1.
\end{equation*}
For a PPM on $k > 2$ patches, the estimate is
\begin{equation*}
L = \alpha_k \, (\mathrm{HP} - 1) + (1 - \alpha_k) \, (\mathrm{MST} - 1),
\end{equation*}
where $\mathrm{HP}$ is the half perimeter of the smallest rectangle
containing the $k$ patches, $\mathrm{MST}$ is the length of a
minimum spanning tree over the $k$ patches, and the blend $\alpha_k$ depends on
$k$.  $L$ is an estimate, not the real length: a patch in the way
can force a detour, and a freed patch can open a shortcut.  The
score of a placement is the sum of $L$ over all PPMs of the
sequence, so a good placement puts patches that interact often
close to each other.

\sys maps in three phases.  First, it builds five starting
placements:
\begin{itemize}
\item One follows the qubit order: number the tiles that may hold
  data patches row by row, and put patch $i$ on the $i$-th one.
\item One ranks the patches by a breadth-first walk of the
  interaction graph~\cite{cuthill1969reducing}, so patches sharing PPMs get nearby ranks, and fills the grid in that order.
\item One lays the ranked patches along a Hilbert
  curve~\cite{hilbert1891curve}, a path that visits the grid tile
  by tile while staying local, so nearby ranks land on nearby tiles.
\item One rounds a spectral layout of the interaction graph onto the grid.
\item One runs an approximate solver for the quadratic assignment problem.
\end{itemize}
Second, \sys improves the placement each strategy produces with small moves,
swapping two patches or moving one patch to a free tile, as long as
the sum of $L$ drops: each strategy ends with its best placement.
Third, the best placement of each strategy goes to the real router
(Section~\ref{sec:routing}): the router routes every PPM of the
sequence on each of these placements, and the placement with the smallest real
total ancilla path length wins.  The router alone is too slow to try
every placement, which is why $L$ does the filtering and the router
only judges the few best placements.  These strategies are heuristics,
but none of them uses randomness: small moves are tried in a fixed
order, so the same input always gives the same placement.

\subsection{Routing}\label{sec:routing}

Every PPM needs an ancilla path, and the router's task is to find a short one.
The ancilla path must avoid the tiles of allocated data patches; every
other tile is free, including the tiles of unallocated or freed
patches.  This is where dynamic allocation helps: a freed tile becomes
usable, so an ancilla path that would detour runs straight and
gets shorter, as in Figure~\ref{fig:overview}(b) and (c).

The router runs a shortest-path search over the free tiles, in
the style of Dijkstra's algorithm~\cite{dijkstra1959note}: for a
PPM on at most nine patches it returns the smallest ancilla path, and
beyond nine patches it runs a fast heuristic.  The exact search
grows exponentially with the number of patches.  We use a threshold
of nine patches to balance ancilla path quality and compile time.

\begin{figure}
  \centering
  \includegraphics[width=\columnwidth]{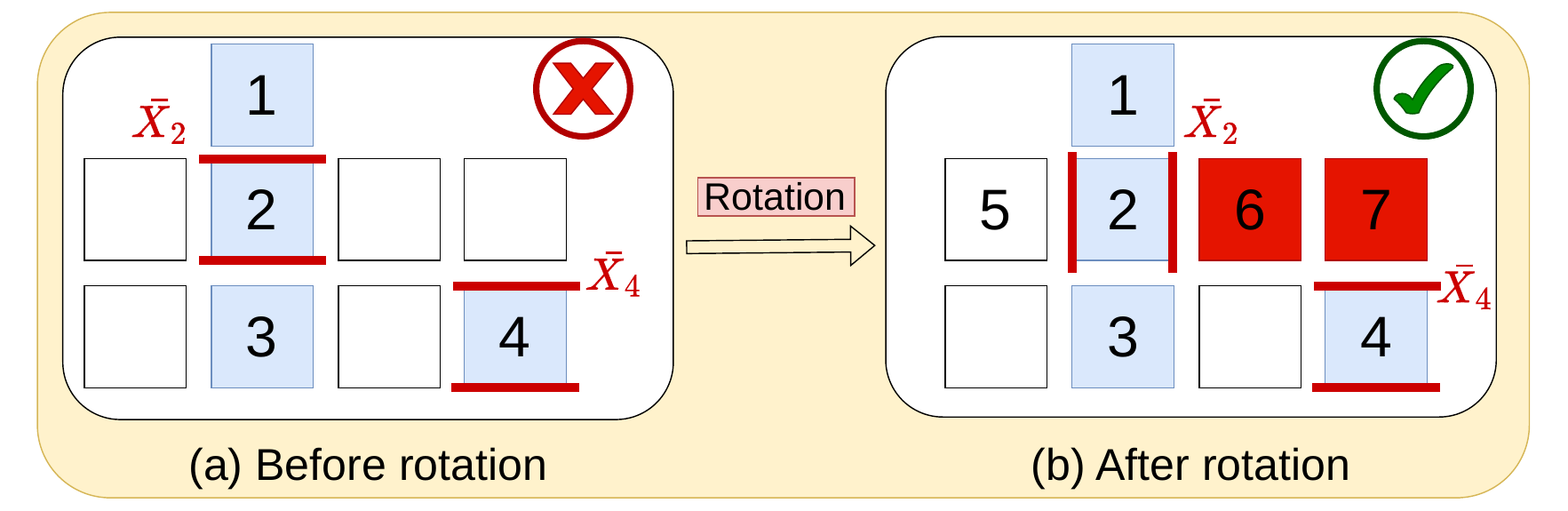}
  \caption{Rotation enables an ancilla path.  (a)~Before
    rotation: the PPM $\bar{X}_2\bar{X}_4$ needs an ancilla path
    between the $\bar{X}$ boundaries of patches 2 and 4, but both
    $\bar{X}$ boundaries of patch 2 touch allocated patches, so no
    ancilla path exists.  (b)~After rotating patch 2 in
    place~\cite{litinski2019game}, $\bar{X}_2$ sits on its left and
    right boundaries, and the ancilla path (tiles 6 and 7) connects
    on the right.}
  \label{fig:rotation}
\end{figure}

As shown in Algorithm~\ref{alg:routing}, the search takes two
inputs.  $F$ collects the free tiles: every tile not held by an
allocated data patch.  $G = \{G_1, \dots, G_k\}$ collects one
group per patch of the PPM: $G_i$ holds the free tiles next to the
boundaries of patch $i$ that support its measured operator.  In
Figure~\ref{fig:rotation}, for example, before the rotation the
$\bar{X}$ boundaries of patch 2 are its top and bottom sides, but
the tiles next to both of them are held by data patches, so $G_2$
is empty.  After the rotation, $\bar{X}_2$ moves to the left
and right sides, and $G_2 = \{5, 6\}$.  The output is an ancilla path:
a connected set of tiles, as small as possible, that touches every
group in $G$.

\begin{algorithm}[t]
\caption{Find Ancilla Path}
\label{alg:routing}
\small
\KwIn{free tiles $F$; groups $G = \{G_1, \dots, G_k\}$, each $G_i$ nonempty}
\KwOut{an ancilla path that touches every group; smallest when $k \le 9$}
\eIf{$k \le 9$}{
  $c[\{i\}, v] \gets 1$ for every group $G_i$ and every tile
    $v \in G_i$; all other entries start at $\infty$\;
  \Repeat{no entry of $c$ changes}{
    \emph{walk:} $c[S, u] \gets \min\{c[S, u],\ c[S, v] + 1\}$
      for every neighbor $u \in F$ of $v$\;
    \emph{merge:} $c[S_1 \cup S_2, v] \gets
      \min\{c[S_1 \cup S_2, v],\ c[S_1, v] + c[S_2, v] - 1\}$\;
  }
  \Return the smallest entry $c[\{1, \dots, k\}, v]$; its tree
    is rebuilt from the recorded moves\;
}{
  $T \gets$ one tile of the group with the fewest tiles\;
  \While{some group is not attached}{
    run a BFS from all tiles of $T$ over the free tiles\;
    add the first-reached tile of an unattached group, and the
      shortest path to it, into $T$\;
  }
  \Return $T$\;
}
\end{algorithm}

Case 1: $k \le 9$.  The finished ancilla path must touch all $k$
groups.  On the way there, the algorithm builds unfinished pieces
of it, which we call \emph{partial trees}: a partial tree is a
connected set of tiles that touches one or more of the groups, but
not yet all of them.  Take Figure~\ref{fig:corridor-search}(a).
Every tile of every group starts as a one-tile partial tree: in the
first panel, tile 9 is a partial tree for patch 1's group, tile 8
for patch 3's, and tile 4 for patch 2's, each drawn in its own
color.  Two moves build bigger trees.  A \emph{walk} grows a
partial tree into a neighboring free tile, adding one tile: the red
tree walks from 9 into 10, and later from 10 into 11.  A
\emph{merge} glues two partial trees that stand on a common tile
into one: in the second panel, the purple tree and the green tree
both reach tile 5 and merge, and the merged tree now touches the
groups of patches 3 and 2.  In the last panel, the merged tree and
the red tree meet at tile 7; merging them gives a tree that touches
all three groups, and that tree is the ancilla path.  Behind the
pictures the algorithm keeps a table $c$: the entry $c[S, v]$ is
the size of the smallest partial tree found so far that touches
every group in the set $S$ and stands on tile $v$.  In
Figure~\ref{fig:corridor-search}(a), the red tree of the second
panel is the entry $c[\{1\}, 11] = 3$: a three-tile tree that
touches patch 1's group and stands on tile 11.  The two moves update the table.  A walk from tile $v$ into a
neighboring free tile $u$ adds one tile:
$c[S, u] \gets c[S, v] + 1$.  A merge of two trees on the same
tile $v$ adds their sizes:
$c[S_1 \cup S_2, v] \gets c[S_1, v] + c[S_2, v] - 1$, minus one
because both trees count the shared tile.  The moves run again and
again, and an entry drops whenever a walk or a merge finds a
smaller tree for it.  When no walk or merge changes an entry, the
table is final, and the smallest entry that touches all $k$ groups
wins.

Case 2: $k > 9$. \sys builds five candidate trees using
different orders for connecting the groups.
Algorithm~\ref{alg:routing} shows the nearest-first order. Each tree starts
at a tile of its first group. To attach a new group, a
breadth-first search starts from all tiles of the current
tree and finds a shortest path through the free tiles to
that group. Groups already touched by the tree are skipped.  The five
orders are:
\begin{itemize}
\item \textbf{Nearest first.} The tree starts from the group
  with the fewest tiles. Each search then connects the nearest
  group not yet touched by the tree
  (Figure~\ref{fig:corridor-search}(b)).
\item \textbf{PPM order.} The groups follow the order in which
  their patches appear in the PPM, from $G_1$ to $G_k$.
\item \textbf{Reverse order.} The groups follow the PPM order
  backwards, from $G_k$ to $G_1$.
\item \textbf{Smallest group first.} The groups are sorted by
  the number of tiles they contain. Groups with fewer tiles
  are connected first.
\item \textbf{Grid order.} Scan the grid by column and then
  by row. Connect groups in the order in which their tiles
  are first encountered.
\end{itemize}

Different orders can produce different trees. \sys discards
duplicate candidates and keeps the tree with the fewest tiles.

Case 1 is the Steiner-tree dynamic program of Erickson, Monma and Veinott~\cite{erickson1987sendsplit}: within nine patches, it always
finds the smallest ancilla path.  Case 2 is a heuristic: fast at any size, but
the tree is not always the smallest.

Routing has one more tool: rotating a patch in place.  Rotation
enables routing to a patch whose legal sides all touch allocated
patches, and it can also shorten ancilla paths.  A patch has two kinds of boundaries: one pair
supports the logical $\bar{X}$ and the other pair the logical
$\bar{Z}$, as in Figure~\ref{fig:rotation}.  The two pairs
split the four sides: if $\bar{X}$ takes the top and bottom sides,
$\bar{Z}$ takes the left and right; if $\bar{X}$ takes the left and
right sides, $\bar{Z}$ takes the top and bottom.  The
ancilla path must reach the measured operator, so the PPM fixes which
sides the ancilla path can attach to.  When the free tiles touch
neither of the two legal sides, no ancilla path exists, as for patch 2
in Figure~\ref{fig:rotation}(a): both of its $\bar{X}$ boundaries
touch allocated patches.  \sys must then rotate the patch in
place, which changes the orientation of its logical
operators~\cite{litinski2019game}.  After the rotation \sys routes
again.  With
\sys's own mapping this blocking does not happen, because every patch
keeps free tiles on all four sides (Section~\ref{sec:mapping}); the
fallback keeps \sys compatible with denser mappings from other
tools.  Besides
unblocking a patch, rotation can also save ancilla path length.
Rotating a patch adds operations, so \sys weighs this extra cost
against the benefit of a shorter ancilla path.  When rotation is
enabled to shorten paths, \sys scores each
option, rotated or not, as $\ell + \lambda \, r$.  Here, $\ell$ is
the option's ancilla path length, $r$ is 1 if the option rotates
any patch and 0 otherwise, and $\lambda$ is the rotation penalty,
expressed in ancilla path tiles.  \sys chooses the lowest score,
breaking ties in favor of the shorter ancilla path.

\section{Lowering PPM Sequences to Physical Circuits}\label{sec:framework}
\begin{table}
\caption{The four cases and their stabilizer constructions.}
\label{tab:four-cases}
\centering
\footnotesize
\renewcommand{\arraystretch}{0.92}
\begin{tabular}{ccl}
\toprule
bases & shapes & stabilizer construction \\
\midrule
same      & same      & pure seam (Figure~\ref{fig:four-constructions}(a)) \\
different & same      & mixed seam (Figure~\ref{fig:four-constructions}(b)) \\
same      & different & stretched pure seam (Figure~\ref{fig:four-constructions}(c)) \\
different & different & stretched mixed seam (Figure~\ref{fig:four-constructions}(d)) \\
\bottomrule
\end{tabular}
\end{table}

Section~\ref{sec:mapping-routing} left every PPM with placed patches and an
ancilla path.  This section turns the sequence into a physical
circuit.  \sys processes the PPMs in order, and each PPM asks for
the same two steps: attach every measured patch to the ancilla path
(Section~\ref{sec:seams}), and construct the stabilizers inside
the ancilla path (Section~\ref{sec:rules}).  The two steps yield the
stabilizer construction of the merged patch, satisfying the three
requirements of Section~\ref{sec:construction-bg}; the construction is rule-based
and runs in time linear in the ancilla path size.

The step from the constructed stabilizers to the circuit is
simple; the hard part is constructing the stabilizers.  A stim
circuit is determined by the stabilizers being measured and their
syndrome extraction schedules.  \sys then builds the circuit from
the constructed stabilizers and these schedules.  LightStim, an
existing tool, annotates the detectors and
observables~\cite{lightstim2026autodem}.

\begin{figure*}
  \centering
  \begin{minipage}[b]{0.225\textwidth}\centering
    \includegraphics[width=\linewidth]{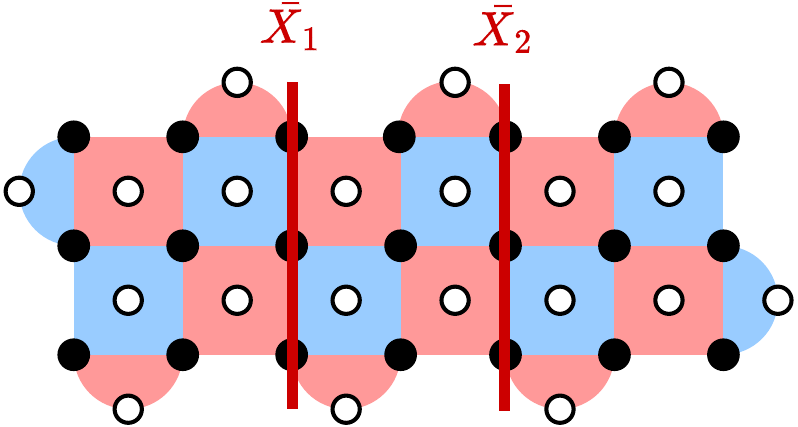}\\ (a)~Pure seam
  \end{minipage}\hfill
  \begin{minipage}[b]{0.225\textwidth}\centering
    \includegraphics[width=\linewidth]{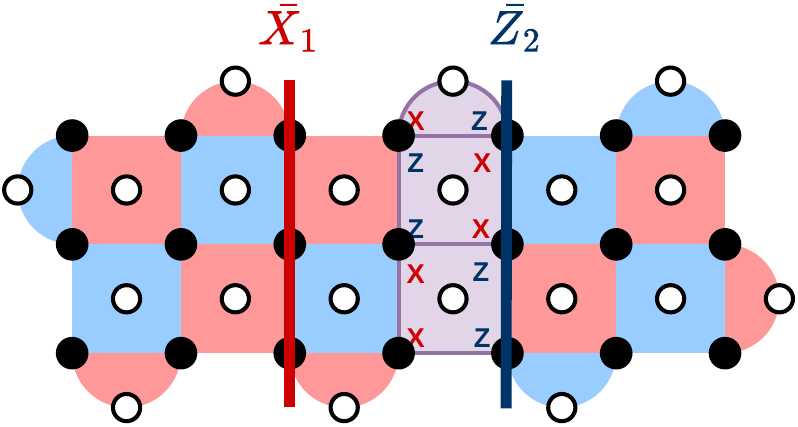}\\ (b)~Mixed seam
  \end{minipage}\hfill
  \begin{minipage}[b]{0.225\textwidth}\centering
    \includegraphics[width=\linewidth]{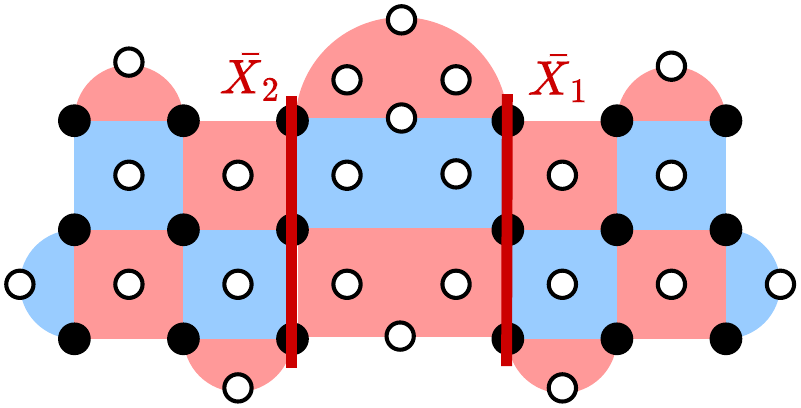}\\ (c)~Stretched pure seam
  \end{minipage}\hfill
  \begin{minipage}[b]{0.225\textwidth}\centering
    \includegraphics[width=\linewidth]{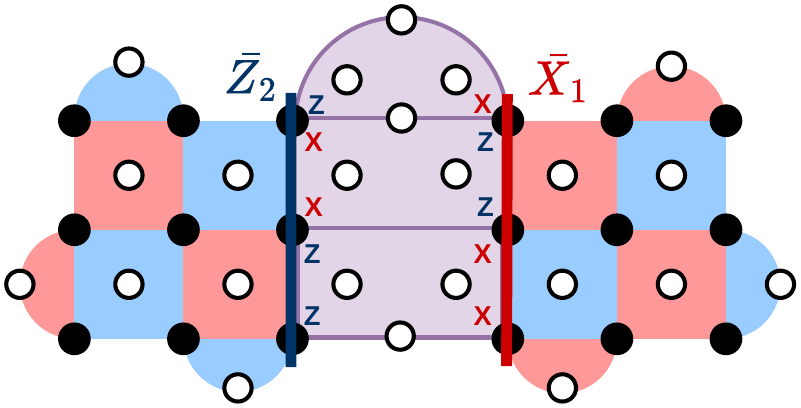}\\ (d)~Stretched mixed seam
  \end{minipage}
  \caption{The four seam constructions.  In (c) and (d), patches 1
    and 2 have different shapes.}
  \label{fig:four-constructions}
\end{figure*}

\begin{figure*}
  \centering
  \includegraphics[width=\textwidth]{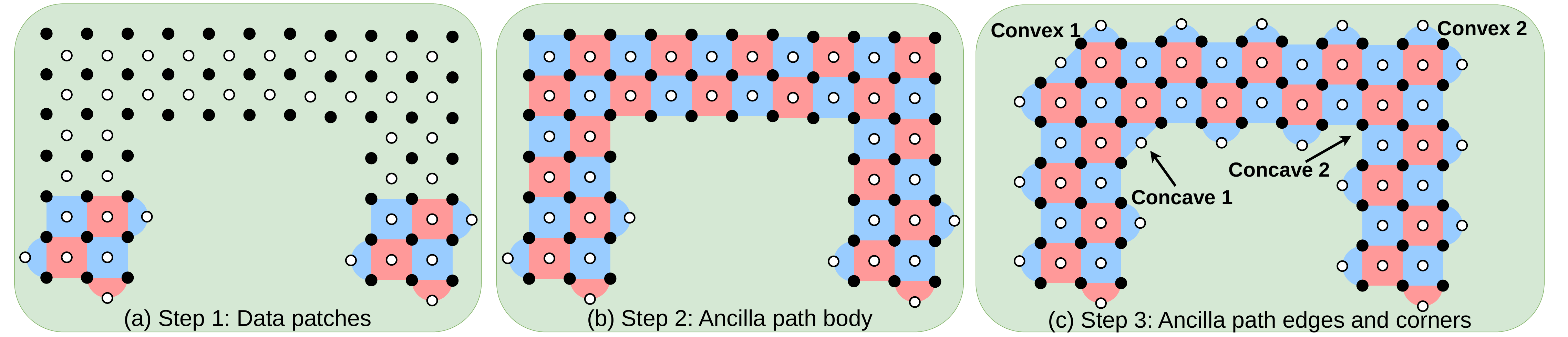}
  \caption{The construction rules of the ancilla path, applied in
    three steps in order.  (a)~Step~1, data patches: each patch keeps its
    ordinary stabilizers, minus the weight-2 stabilizer on the
    boundary that meets the ancilla path.  (b)~Step~2, ancilla path
    body: the inside of the ancilla path is filled with weight-4
    stabilizers.  (c)~Step~3, ancilla path edges and corners:
    boundary stabilizers at every other gap along the edges, and at
    each corner by its corner rule; Convex~1/2 and Concave~1/2 mark
    the two cases of each corner rule.}
  \label{fig:rules}
\end{figure*}

We give general stabilizer construction rules.  Prior work shows
such constructions as worked examples.  Litinski draws multi-patch
merges in Figures~37 and~41 of~\cite{litinski2019game}, Kishony
and Fowler draw two more in Figures~2 and~3
of~\cite{kishony2026surface}, and mixed seams appear
in~\cite{geher2024hadamard,chamberland2022twistfree}.  The
construction below covers every legal ancilla path the router
returns.  An ancilla path is legal when it is a
tree of free tiles, and every measured patch has an ancilla path tile
next to a boundary supporting its measured logical operator
(Figure~\ref{fig:rotation}(b)).

\subsection{The Stabilizers Between a Measured Patch and the Ancilla Path}\label{sec:seams}

A patch has two properties.  The first is the \emph{measurement
basis}: whether the PPM measures the patch in the $X$ basis or the
$Z$ basis.  The second is the \emph{shape}: where the weight-2 stabilizers
sit on the boundary.  Only two sets of positions are possible,
and the shape says which one the patch uses.  The shape looks
only at the positions, not at whether the stabilizers are $X$ or
$Z$.

The ancilla path tile that attaches to a patch also has a measurement
basis and a shape: in Figure~\ref{fig:rotation}(b), tile 6 is
that tile for
patch 2, and tile 7 for patch 4.  One ancilla path has one basis and
one shape for all its tiles.  \sys sets the basis to the
majority basis of the measured patches, with ties broken toward
$X$; the shape is whichever of the two lets the construction
succeed.  

What matters is whether the measured patch and the ancilla path
tile have the same or different measurement bases and shapes.
These combinations give four cases, listed in
Table~\ref{tab:four-cases}.  Each case has its own construction of
the stabilizers between the patch and the ancilla path.  We call
these stabilizers the \emph{seam}.  The four constructions are
shown in Figure~\ref{fig:four-constructions}.  The stretched
constructions follow~\cite{litinski2019game,kishony2026surface}.
The stabilizers of a mixed seam have both $X$ and $Z$
support~\cite{geher2024hadamard,chamberland2022twistfree}.

\subsection{Construction Rules for the Ancilla Path}\label{sec:rules}

Section~\ref{sec:seams} handled where the ancilla path meets each
patch. This section describes how to construct the stabilizers
for the ancilla path. Figure~\ref{fig:rules} shows an example of
this construction on a bent ancilla path in three steps.
The general stabilizer construction for the ancilla path is built
by the following five rules:

\noindent\textbf{Data patches.}  Every data patch keeps its ordinary
  stabilizers.  Only the side that meets the ancilla path changes: the
  weight-2 stabilizers on that boundary are removed
  (Figure~\ref{fig:rules}(a)).

\noindent\textbf{Ancilla path body.}  The inside of the ancilla path is
  filled with ordinary weight-4 stabilizers
  (Figure~\ref{fig:rules}(b)).

\noindent\textbf{Ancilla path edges.}  Along the edges of the
  ancilla path, boundary stabilizers are added at every other gap:
  add one, skip one, add one, starting next to the last weight-2
  stabilizer the patch keeps (Figure~\ref{fig:rules}(c)).  A
  \emph{gap} is the slot between two neighboring data qubits
  along the boundary, and each gap holds at most one boundary
  stabilizer.  Each added stabilizer is weight-2; the corners are
  special, and the next two rules handle them.

\noindent\textbf{Concave inward corners.}  The inner corner
  follows the same spacing: if the corner position is two gaps
  away from its neighboring boundary stabilizer, a weight-3
  stabilizer is added (concave inward corner 1 in
  Figure~\ref{fig:rules}(c)); if it is one gap away, nothing is
  added (concave inward corner 2 in Figure~\ref{fig:rules}(c)).

\noindent\textbf{Convex outer corners.}  The outer corner of a bend
  is also checked by spacing: two gaps away, nothing changes
  (convex outer corner 2 in Figure~\ref{fig:rules}(c)); one gap
  away, the corner data qubit is removed, and the corner stabilizer
  is cut from weight-4 to weight-3 (convex outer corner 1 in
  Figure~\ref{fig:rules}(c)).

\FloatBarrier
\section{Evaluation}\label{sec:evaluation}

\begin{figure}[t]
  \centering
  \includegraphics[width=\columnwidth]{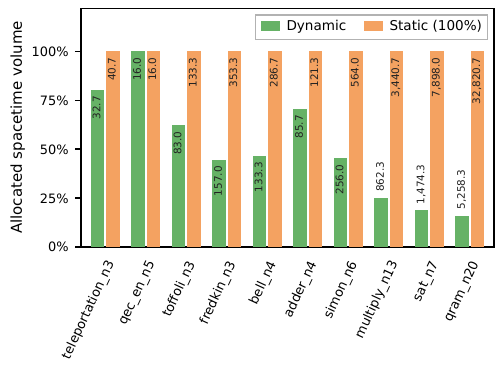}
  \caption{Allocated spacetime volume of Dynamic and Static for ten
    programs at $d=3$.}
  \label{fig:occupancy-ten}
\end{figure}

\begin{table*}[h]
  \centering
  \caption{Comparison to two baselines on non-Clifford programs.
    TQEC$_1$ = TopoLS + \texttt{tqec}; TQEC$_2$ =
    topologiq + \texttt{tqec}; $\times$: no valid value.}
  \label{tab:comparison}
  \fontsize{8}{9.4}\selectfont
  \setlength{\tabcolsep}{2pt}
  \setlength{\arrayrulewidth}{0.35pt}
  \setlength{\doublerulesep}{1pt}
  \renewcommand{\arraystretch}{1.12}
  \begin{tabularx}{\textwidth}{|>{\raggedright\arraybackslash}p{88pt}|>{\centering\arraybackslash}p{12pt}||*{3}{>{\centering\arraybackslash}X|}|*{3}{>{\centering\arraybackslash}X|}|*{3}{>{\centering\arraybackslash}X|}|*{3}{>{\centering\arraybackslash}X|}}
    \hline
    & & \multicolumn{3}{c||}{allocated volume (blocks)}
      & \multicolumn{3}{c||}{qubit-cycles}
      & \multicolumn{3}{c||}{compile time (s)}
      & \multicolumn{3}{c|}{LER} \\
    \hline
    program & $d$ & TQEC$_1$ & TQEC$_2$ & Ours & TQEC$_1$ & TQEC$_2$ & Ours & TQEC$_1$ & TQEC$_2$ & Ours & TQEC$_1$ & TQEC$_2$ & Ours \\
    \hline\hline
    teleportation\_n3 & 3 & 48 & 26 & 32.7 & 3162 & 1629 & 1712 & 27.8 & 15.8 & 0.41 & 0.029 & 0.021 & 0.019 \\
    \hline
    teleportation\_n3 & 5 & 48 & 26 & 27.6 & 13718 & 7175 & 6770 & 39.8 & 20.2 & 1.29 & 0.0031 & 0.0024 & 0.0025 \\
    \hline
    toffoli\_n3 & 5 & 123 & 96 & 69.4 & 34290 & 26565 & 17174 & 115 & 96.1 & 4.94 & 0.0028 & 0.0077 & 0.0098 \\
    \hline
    fredkin\_n3 & 3 & 122 & 111 & 157 & 7662 & 7110 & 8202 & 237 & 36.4 & 2.06 & 0.16 & 0.17 & 0.12 \\
    \hline
    fredkin\_n3 & 5 & 122 & 96 & 135 & 33730 & 26885 & 33000 & 315 & 172 & 11.2 & 0.021 & 0.02 & 0.019 \\
    \hline
    adder\_n4 & 3 & 165 & 147 & 85.7 & 10338 & 9372 & 4566 & 315 & 122 & 1.03 & 0.23 & 0.19 & 0.019 \\
    \hline
    adder\_n4 & 5 & 165 & 147 & 71.4 & 45590 & 41000 & 17742 & 356 & 153 & 5.89 & 0.038 & 0.022 & 0.0026 \\
    \hline
    qec\_en\_n5 & 3 & $\times$ & 142 & 16 & $\times$ & 9192 & 775 & $\times$ & 136 & 0.27 & $\times$ & 0.18 & 0.017 \\
    \hline
    simon\_n6 & 5 & $\times$ & 254 & 215 & $\times$ & 70700 & 52717 & $\times$ & 333 & 26.2 & $\times$ & 0.016 & 0.049 \\
    \hline
    toffoli\_n3 & 3 & 123 & $\times$ & 83 & 7816 & $\times$ & 4386 & 93.8 & $\times$ & 0.98 & 0.025 & $\times$ & 0.062 \\
    \hline
    qec\_en\_n5 & 5 & $\times$ & $\times$ & 13.2 & $\times$ & $\times$ & 3087 & $\times$ & $\times$ & 0.69 & $\times$ & $\times$ & 0.0027 \\
    \hline
    bell\_n4 & 3 & $\times$ & $\times$ & 133 & $\times$ & $\times$ & 7025 & $\times$ & $\times$ & 1.71 & $\times$ & $\times$ & 0.12 \\
    \hline
    bell\_n4 & 5 & $\times$ & $\times$ & 112 & $\times$ & $\times$ & 27643 & $\times$ & $\times$ & 9.39 & $\times$ & $\times$ & 0.019 \\
    \hline
    simon\_n6 & 3 & $\times$ & $\times$ & 256 & $\times$ & $\times$ & 13403 & $\times$ & $\times$ & 3.72 & $\times$ & $\times$ & 0.27 \\
    \hline
    multiply\_n13 & 3 & 1425 & $\times$ & 862 & 87381 & $\times$ & 44306 & 3047 & $\times$ & 38.8 & $\times$ & $\times$ & 0.49 \\
    \hline
    multiply\_n13 & 5 & 1425 & 912 & 718 & 388375 & 252615 & 174018 & 3489 & 1589 & 322 & $\times$ & $\times$ & 0.1 \\
    \hline
    sat\_n7 & 3 & $\times$ & $\times$ & 1474 & $\times$ & $\times$ & 78203 & $\times$ & $\times$ & 140 & $\times$ & $\times$ & 0.49 \\
    \hline
    qram\_n20 & 3 & $\times$ & $\times$ & 5258 & $\times$ & $\times$ & 270106 & $\times$ & $\times$ & 1771 & $\times$ & $\times$ & 0.98 \\
    \hline
    \multicolumn{2}{|l||}{\begin{tabular}[c]{@{}l@{}}Ours vs.\ TQEC$_1$,\\geometric mean\end{tabular}} & \multicolumn{3}{c||}{$-34\%$} & \multicolumn{3}{c||}{$-44\%$} & \multicolumn{3}{c||}{$-98\%$} & \multicolumn{3}{c|}{$-40\%$} \\
    \hline
    \multicolumn{2}{|l||}{\begin{tabular}[c]{@{}l@{}}Ours vs.\ TQEC$_2$,\\geometric mean\end{tabular}} & \multicolumn{3}{c||}{$-27\%$} & \multicolumn{3}{c||}{$-38\%$} & \multicolumn{3}{c||}{$-96\%$} & \multicolumn{3}{c|}{$-48\%$} \\
    \hline
  \end{tabularx}
\end{table*}

In this section, we evaluate \sys from different perspectives.
Section~\ref{sec:methodology} describes the setup.
Section~\ref{sec:comparison} compares \sys with two baselines.  Section~\ref{sec:ablations} ablates each rule.
Section~\ref{sec:faulttol} checks fault tolerance.
Section~\ref{sec:validation} tests the LER formula of prior work.

\subsection{Methodology}\label{sec:methodology}

\vspace{2pt}\noindent\textbf{\textit{Benchmarks.}}  The suite has ten
non-Clifford programs from QASMBench, from 3 to 20
qubits~\cite{li2023qasmbench}.  Table~\ref{tab:comparison} contains
18 rows, each corresponding to a program at $d = 3$ or $d = 5$.

\vspace{2pt}\noindent\textbf{\textit{Magic-state proxies.}}  A stim
circuit supports only Clifford operations~\cite{gidney2021stim}, so
we use a $|+\rangle$ state as a proxy for every $|T\rangle$ state.  The
proxy changes nothing the compiler sees: the PPM sequence, the
merges and readout of every T gadget, and every mapping and routing
decision stay as they are, while the circuit becomes Clifford and
samples exactly.  Prior work makes the same proxy with an
$|i\rangle$ state, turning each T gate into an S
gate~\cite{gidney2024cultivation,gidney2023hook,gidney2024inplacey,hao2025breakeven};
we use $|+\rangle$ because \texttt{tqec} has no $Y$-basis block, so
the two baselines are proxied using a $|+\rangle$ state.

\vspace{2pt}\noindent\textbf{\textit{Configuration.}}  \sys's front end
lowers QASM to a PPM sequence using
NWQEC~\cite{wang2026transpiler,wang2025tableau}, then re-selects the
PPM sequence~\cite{paykin2023pcoast,schmitz2023pcoastext,peres2025pbcweights,sethi2026ppr}.  We use a
circuit-level depolarizing noise model at physical error rate $p$,
including idle locations.  \sys emits stim
circuits~\cite{gidney2021stim}, decoded with
\texttt{PyMatching}~\cite{higgott2025sparseblossom}.  The LER results in Table~\ref{tab:comparison},
Table~\ref{tab:ablation} and Figure~\ref{fig:dscaling}
use $p = 5 \times 10^{-4}$.

\vspace{2pt}\noindent\textbf{\textit{Metrics.}}  \emph{Allocated spacetime
volume}, our headline metric, follows the
spacetime cost of prior work~\cite{litinski2019game}: $s$ tiles
held for $t$ time steps cost $s \cdot t$ blocks, one block being
one tile held for $d$ code cycles.  \emph{Qubit-cycles} measures the same cost at the circuit level:
each physical qubit counts one unit for each code cycle it is held.
\emph{Compile time} is the time
the compiler takes to turn a QASM
program into a circuit.  The baselines run \texttt{tqec} with its
default settings and an empty detector database, that is, a first
compile of each program.  \emph{LER} is the logical error rate.

\subsection{Overall Performance}\label{sec:comparison}

We run the same quantum programs through three toolchains,
TQEC$_1$ (TopoLS + \texttt{tqec}~\cite{topols2026,tqec2026joss}),
TQEC$_2$ (topologiq + \texttt{tqec}~\cite{bolanos2025topologiq,tqec2026joss})
and \sys, and report the results in Table~\ref{tab:comparison}.
All three compile the same QASM program to a stim circuit, and the
LER comes from sampling the compiled circuit.

\begingroup
\setlength{\emergencystretch}{3em}
\vspace{2pt}\noindent\textbf{\textit{Metric comparison.}}
Compared with TQEC$_1$, \sys reduces allocated volume by $34\%$,
qubit-cycles by $44\%$, compile time by $98\%$ and LER by $40\%$.
Compared with TQEC$_2$, \sys reduces allocated volume by $27\%$,
qubit-cycles by $38\%$, compile time by $96\%$ and LER by $48\%$.
These reductions are computed from geometric means over rows
where both toolchains report the corresponding metric.
\par\endgroup

\begin{table*}[tb]
  \centering
  \caption{Ablations at $d = 3$. $\Delta\%$ denotes the percentage
    change relative to the full pipeline.}
  \label{tab:ablation}
  \fontsize{9}{10.5}\selectfont
  \setlength{\tabcolsep}{4pt}
  \setlength{\arrayrulewidth}{0.35pt}
  \setlength{\doublerulesep}{1pt}
  \renewcommand{\arraystretch}{1.12}
  \begin{tabularx}{\textwidth}{|>{\hsize=1.7\hsize\linewidth=\hsize\raggedright\arraybackslash}X|>{\hsize=1.15\hsize\linewidth=\hsize\centering\arraybackslash}X|>{\hsize=1.0\hsize\linewidth=\hsize\centering\arraybackslash}X|>{\hsize=1.0\hsize\linewidth=\hsize\centering\arraybackslash}X|>{\hsize=0.65\hsize\linewidth=\hsize\centering\arraybackslash}X|>{\hsize=0.5\hsize\linewidth=\hsize\centering\arraybackslash}X|}
    \hline
    configuration & allocated volume $\Delta\%$ & qubit-cycles $\Delta\%$
      & compile time $\Delta\%$ & LER $\Delta\%$ & fails \\
    \hline\hline
    w/o first-use initialization & $+85.3$ & $+68.6$ & $+36.3$ & $+25.7$ & 0 \\
    \hline
    w/o last-use freeing & $+84.9$ & $+78.8$ & $+102.1$ & $+34.0$ & 0 \\
    \hline
    w/o mapping & $+25.0$ & $+30.2$ & $+13.5$ & $+10.9$ & 0 \\
    \hline
  \end{tabularx}
\end{table*}

\vspace{2pt}\noindent\textbf{\textit{Why \sys wins.}}  \sys wins
on allocated volume, qubit-cycles and LER.  For TQEC$_1$, every
data patch holds its tile from the start of the
program to the end, which has two consequences: the patch is
exposed to idle noise throughout its live range, and later ancilla
paths have to detour around the held tiles and grow longer.  \sys
allocates a patch at its first use and frees it at its last use.
For TQEC$_2$, \sys has better heuristics that increase the
locality of connected patches: it places the patches a PPM
measures together closer, so their ancilla paths are shorter.

\sys also compiles faster.  It reduces compile time by $98\%$ and
$96\%$ in geometric mean.  \sys uses fast heuristics for mapping and
for routing larger PPMs, and its stabilizer construction rules run in
linear time.  Both baselines spend most of their time on circuit
generation in \texttt{tqec}.  TQEC$_1$ is the slower of the two
because TopoLS~\cite{topols2026} searches for a layout with Monte
Carlo tree search, which takes from 9 seconds to 27 minutes on our
programs; the greedy layout of topologiq~\cite{bolanos2025topologiq}
takes under a minute.

\vspace{2pt}\noindent\textbf{\textit{\sys compiles what the
baselines cannot.}} Table~\ref{tab:comparison} has 18 rows.
TQEC$_1$ has no valid entry on 8 of them, and TQEC$_2$ has no
valid entry on 8 of them, while \sys delivers all of them.
TQEC$_1$ misses rows for three reasons: TopoLS finds no layout
that realizes the program, \texttt{tqec} fails to compile the
layout TopoLS finds, or the compiled circuit leaves observables that
are not deterministic.  TQEC$_2$ misses rows for two reasons:
\texttt{tqec} rejects topologiq's layout, or it compiles the
layout into a circuit whose detectors or observables are not
deterministic.

Where a baseline does deliver a circuit, its LER can still be
missing: this occurs on 2 rows of TQEC$_1$ and 1 row of TQEC$_2$.
The TQEC$_1$ circuits contain detectors that are random at $p = 0$.
For the TQEC$_2$ circuit, \texttt{stim} fails to decompose the
error model for \texttt{PyMatching}.

\subsection{Ablations}\label{sec:ablations}

We run an ablation study: each ``w/o'' configuration in
Table~\ref{tab:ablation} removes one rule of
Section~\ref{sec:lifetime} or the mapper of
Section~\ref{sec:mapping}.

\vspace{2pt}\noindent\textbf{\textit{Spacetime savings from
dynamic allocation.}}
Figure~\ref{fig:occupancy-ten} compares the allocated spacetime
volume of Dynamic and Static for ten programs at $d=3$.
Static disables both first-use initialization and last-use freeing.
Dynamic allocation reduces allocated spacetime volume by up to
$84.0\%$, demonstrating its effectiveness in reducing spacetime costs.

The savings are more pronounced for the larger programs in our
benchmark suite. The seven smaller programs achieve reductions
of $0\%$ to $55.6\%$ in allocated spacetime volume.
The larger programs \texttt{sat\_n7}, \texttt{multiply\_n13},
and \texttt{qram\_n20} achieve reductions of $74.9\%$ to $84.0\%$.
These results highlight the potential of our compiler to
substantially reduce the spacetime overhead of large-scale FTQC.

\vspace{2pt}\noindent\textbf{\textit{The goal is LER; the rest are tools.}}
A fault-tolerant program aims for a low LER within an acceptable
resource budget, so the number to minimize is the LER.  \sys
lowers the LER by cutting allocated volume and qubit-cycles.  In
Table~\ref{tab:ablation}, removing first-use initialization raises
the allocated volume by $85\%$ and the LER by $26\%$; removing
last-use freeing raises them by $85\%$ and $34\%$; removing the
mapper raises them by $25\%$ and $11\%$.  Each component lowers
the LER, and that is what moves a fault-tolerant program toward
its goal.

\vspace{2pt}\noindent\textbf{\textit{Why each component helps.}}
Without first-use initialization, patches are allocated before
they are needed. This extends their live ranges and exposes
them to idle noise before their first use. Their occupied tiles
can also force longer ancilla paths.

Without last-use freeing, scratch patches remain allocated
after their last use. This adds idle rounds and prevents later
ancilla paths from reusing their tiles. The longer live ranges
and possible detours increase spacetime volume and exposure
to noise.

Without the mapper, patches are placed in row-major order.
This placement ignores which patches are measured together
and can require longer ancilla paths. Longer paths occupy more
tiles and introduce more noisy operations. Every program still
compiles under this placement, but \texttt{qram\_n20} needs $55\%$
more allocated volume.

\subsection{Fault Tolerance}\label{sec:faulttol}

\vspace{2pt}\noindent\textbf{\textit{Verification.}}  We use three
checks to test whether the circuits are fault tolerant and measure
the intended PPMs.
First, the shortest graphlike logical error has weight $d$
on all possible construction cases.
Second, no detector is triggered at $p = 0$.
Third, the sampled output bits obey the same parity constraints
as a logical-level simulation of the same program.

\begin{figure}[t]
  \centering
  \includegraphics[width=\columnwidth]{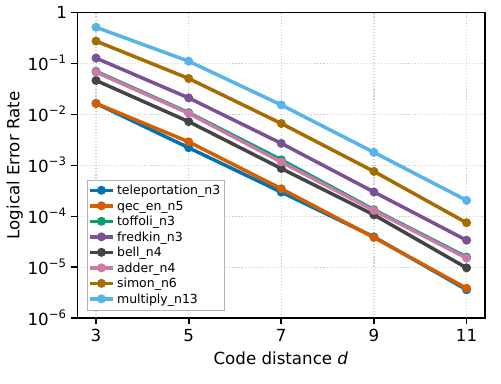}
  \caption{LER of the \sys-compiled circuits of eight programs
    against code distance.}
  \label{fig:dscaling}
\end{figure}

\vspace{2pt}\noindent\textbf{\textit{Distance scaling.}}  \sys keeps the
exponential suppression in distance.  Figure~\ref{fig:dscaling} shows the LER of eight programs at
$d = 3, 5, 7, 9, 11$.  As the figure shows, the LER falls
exponentially with $d$, so the compiled circuits are fault
tolerant: every step in $d$ suppresses the LER by a constant
factor, as the surface code promises.

\subsection{Validating the LER Formula}\label{sec:validation}

The composition formula~\cite{o3ls2026}
predicts the total LER as the sum over layers
\begin{equation*}
p_{\text{total}} \approx \sum_{t=1}^{T}
\bigl(1-(1-P_{\text{PPM}}^{(t)})
(1-P_{\text{PR}}^{(t)})
(1-P_{\text{idle}}^{(t)})\bigr).
\end{equation*}
Our compiled circuits contain no patch rotations, so
$P_{\text{PR}}^{(t)} = 0$ throughout.  For each benchmark,
$p_{\text{meas}}$ is the LER sampled from the compiled circuit,
and $p_{\text{pred}}$ is the value the formula gives when each
layer's PPM and idle terms are simulated in isolation.  The deviation
between them is
$|p_{\text{pred}} - p_{\text{meas}}| / p_{\text{meas}}$, reported
in Table~\ref{tab:formula}.

\begin{table}[tb]
  \centering
  \caption{Median deviation of the formula's prediction from the
    measured LER.}
  \label{tab:formula}
  \fontsize{9}{10.5}\selectfont
  \setlength{\tabcolsep}{4pt}
  \setlength{\arrayrulewidth}{0.35pt}
  \setlength{\doublerulesep}{1pt}
  \renewcommand{\arraystretch}{1.12}
  \begin{tabularx}{\columnwidth}{|>{\hsize=.55\hsize\linewidth=\hsize\centering\arraybackslash}X|>{\hsize=1.15\hsize\linewidth=\hsize\centering\arraybackslash}X|>{\hsize=1.3\hsize\linewidth=\hsize\centering\arraybackslash}X|>{\centering\arraybackslash}X|}
    \hline
    $d$ & $p$ & number of programs & deviation \\
    \hline\hline
    3 & $2 \times 10^{-4}$ & 8 & 33\% \\
    \hline
    3 & $5 \times 10^{-4}$ & 8 & 26\% \\
    \hline
    3 & $1 \times 10^{-3}$ & 8 & 52\% \\
    \hline
    5 & $5 \times 10^{-4}$ & 5 & 34\% \\
    \hline
    5 & $1 \times 10^{-3}$ & 5 & 27\% \\
    \hline
  \end{tabularx}
\end{table}

The result is clear: the formula is not accurate enough.  On half of the benchmarks its prediction deviates by more
than 26 to 52\% at $d = 3$ and by more than 27 to 34\% at
$d = 5$, and on the worst benchmark it deviates by more than
$2\times$.

\section{Related Work}\label{sec:related}

\vspace{2pt}\noindent\textbf{\textit{PPM compilers.}}  Prior compilers
for lattice surgery work on the PPM sequence in two ways.  Some map
and route: they place the patches, then find ancilla paths for each
PPM~\cite{lao2018mapping,beverland2022edpc,molavi2025dascot,hamada2026routing}.
Others optimize the PPM sequence: they reorder commuting PPMs,
run steps in parallel, and shrink the
layout~\cite{litinski2019game,kan2025sparo,o3ls2026,puremagic2025,sethi2026ppr,pflieger2026harvest}.
However, they keep every data patch allocated from start to finish;
they do not free a data patch and reuse its space as an ancilla path.  The
closest work applies first-use timing to magic-state distillation
schedules~\cite{xu2026bicycle}, memory management has
reused ancilla qubits~\cite{ding2020square}, and a
resource-allocating LS compiler frees magic-state factories after
use~\cite{robertson2025resource}; each manages a helper resource,
not the data patches.

\vspace{2pt}\noindent\textbf{\textit{Intermediate representations.}}
Between a quantum program and a physical circuit sits an
intermediate representation, and the field uses two: the PPM
sequence and the ZX diagram (Section~\ref{sec:pipeline}).
Compilers on the PPM side
stop at the
sequence~\cite{wang2026transpiler,wang2025tableau,kan2025sparo,o3ls2026}:
they stay at the logical level and do not emit a physical
circuit for a whole program.  The other route to a
stim circuit runs through the ZX
diagram~\cite{topols2026,tqec2026joss,hao2025breakeven},
and it does not cover the general case: some operators are
unimplemented, so on our benchmarks it fails to produce a
circuit for many programs (Section~\ref{sec:comparison}).  \sys completes
the stack on the PPM side: a quantum program compiles through its
PPM sequence to a runnable stim circuit.

\vspace{2pt}\noindent\textbf{\textit{LER formulas.}}  Existing formulas predict a program's
LER by estimating each part in isolation and combining the
estimates: per-layer
sums~\cite{o3ls2026,kan2025sparo}, block
budgets~\cite{litinski2019game,beverland2022req}, and calibrated
cost models~\cite{huggins2025flasq,hao2025breakeven}.  Among
these, only one checks its estimate against a physical circuit,
and only on two five-qubit programs~\cite{hao2025breakeven}; the
others leave their formulas unchecked.

\section{Conclusion}\label{sec:conclusion}

We presented \sys, a full-stack lattice surgery compiler that
bridges the gap between logical PPM sequences and physical
circuits.  On that basis, \sys allocates patches dynamically: each
patch is allocated at its first use and freed at its last use, so it sits
in noise for less time, and the freed space is reused as ancilla
paths, so the paths get shorter.  Experimental results show that
\sys reaches a 40\% and 48\% lower LER with 34\% and 27\% less
allocated volume than the two prior toolchains, and compiles
programs that they cannot.  We release \sys as open infrastructure, so that
future LS compilers can score their optimizations on
the circuits they actually produce.

\begin{acks}
We thank Xiang Fang for helpful discussions and for guidance on using LightStim~\cite{lightstim2026autodem}.
We are also grateful to Austin Fowler, Jose A. Bolanos and Adrien Suau for their thoughtful feedback on our evaluation.
\end{acks}

\section*{Data Availability}
The source code of \sys is available at
\url{https://github.com/John-YuehanZhang/CircLS}.

\bibliographystyle{ACM-Reference-Format}
\bibliography{references}

@article{preskill2018nisq,
  title   = {Quantum Computing in the {NISQ} era and beyond},
  author  = {John Preskill},
  journal = {Quantum},
  volume  = {2},
  pages   = {79},
  year    = {2018},
  doi     = {10.22331/q-2018-08-06-79}
}

@article{gidneyekera2021factor,
  title   = {How to factor 2048 bit {RSA} integers in 8 hours using 20 million noisy qubits},
  author  = {Craig Gidney and Martin Eker{\aa}},
  journal = {Quantum},
  volume  = {5},
  pages   = {433},
  year    = {2021},
  doi     = {10.22331/q-2021-04-15-433}
}

@article{fowler2012surface,
  title   = {Surface codes: Towards practical large-scale quantum computation},
  author  = {Austin G. Fowler and Matteo Mariantoni and John M. Martinis and Andrew N. Cleland},
  journal = {Physical Review A},
  volume  = {86},
  number  = {3},
  pages   = {032324},
  year    = {2012},
  doi     = {10.1103/PhysRevA.86.032324}
}

@article{dennis2002topological,
  title   = {Topological quantum memory},
  author  = {Eric Dennis and Alexei Kitaev and Andrew Landahl and John Preskill},
  journal = {Journal of Mathematical Physics},
  volume  = {43},
  number  = {9},
  pages   = {4452--4505},
  year    = {2002},
  doi     = {10.1063/1.1499754}
}

@article{ryananderson2021realtime,
  title   = {Realization of real-time fault-tolerant quantum error correction},
  author  = {C. Ryan-Anderson and J. G. Bohnet and K. Lee and D. Gresh and A. Hankin and J. P. Gaebler and D. Francois and A. Chernoguzov and D. Lucchetti and N. C. Brown and T. M. Gatterman and S. K. Halit and K. Gilmore and J. Gerber and B. Neyenhuis and D. Hayes and R. P. Stutz},
  journal = {Physical Review X},
  volume  = {11},
  number  = {4},
  pages   = {041058},
  year    = {2021},
  doi     = {10.1103/PhysRevX.11.041058}
}

@article{krinner2022realizing,
  title   = {Realizing repeated quantum error correction in a distance-three surface code},
  author  = {Sebastian Krinner and Nathan Lacroix and Ants Remm and Agustin {Di Paolo} and Elie Genois and Catherine Leroux and Christoph Hellings and Stefania Lazar and Fran\c{c}ois Swiadek and Johannes Herrmann and Graham J. Norris and Christian Kraglund Andersen and Markus M{\"u}ller and Alexandre Blais and Christopher Eichler and Andreas Wallraff},
  journal = {Nature},
  volume  = {605},
  number  = {7911},
  pages   = {669--674},
  year    = {2022},
  doi     = {10.1038/s41586-022-04566-8}
}

@article{google2023suppressing,
  title   = {Suppressing quantum errors by scaling a surface code logical qubit},
  author  = {{Google Quantum AI}},
  journal = {Nature},
  volume  = {614},
  number  = {7949},
  pages   = {676--681},
  year    = {2023},
  doi     = {10.1038/s41586-022-05434-1}
}

@article{acharya2025belowthreshold,
  title   = {Quantum error correction below the surface code threshold},
  author  = {{Google Quantum AI and Collaborators}},
  journal = {Nature},
  volume  = {638},
  number  = {8052},
  pages   = {920--926},
  year    = {2025},
  doi     = {10.1038/s41586-024-08449-y},
  note    = {arXiv:2408.13687}
}

@article{bluvstein2024logical,
  title   = {Logical quantum processor based on reconfigurable atom arrays},
  author  = {Dolev Bluvstein and Simon J. Evered and Alexandra A. Geim and Sophie H. Li and Hengyun Zhou and Tom Manovitz and Sepehr Ebadi and Madelyn Cain and Marcin Kalinowski and Dominik Hangleiter and J. Pablo {Bonilla Ataides} and Nishad Maskara and Iris Cong and Xun Gao and Pedro {Sales Rodriguez} and Thomas Karolyshyn and Giulia Semeghini and Michael J. Gullans and Markus Greiner and Vladan Vuleti{\'c} and Mikhail D. Lukin},
  journal = {Nature},
  volume  = {626},
  number  = {7997},
  pages   = {58--65},
  year    = {2024},
  doi     = {10.1038/s41586-023-06927-3},
}

@article{erhard2021entangling,
  title   = {Entangling logical qubits with lattice surgery},
  author  = {Alexander Erhard and Hendrik {Poulsen Nautrup} and Michael Meth and Lukas Postler and Roman Stricker and Martin Stadler and Vlad Negnevitsky and Martin Ringbauer and Philipp Schindler and Hans J. Briegel and Rainer Blatt and Nicolai Friis and Thomas Monz},
  journal = {Nature},
  volume  = {589},
  number  = {7841},
  pages   = {220--224},
  year    = {2021},
  doi     = {10.1038/s41586-020-03079-6},
}

@article{besedin2026lattice,
  title   = {Lattice surgery realized on two distance-three repetition codes with superconducting qubits},
  author  = {Ilya Besedin and Michael Kerschbaum and Jonathan Knoll and Ian Hesner and Lukas B{\"o}deker and Luis Colmenarez and Luca Hofele and Nathan Lacroix and Christoph Hellings and Fran\c{c}ois Swiadek and Alexander Flasby and Mohsen {Bahrami Panah} and Dante {Colao Zanuz} and Markus M{\"u}ller and Andreas Wallraff},
  journal = {Nature Physics},
  volume  = {22},
  number  = {2},
  pages   = {189--194},
  year    = {2026},
  doi     = {10.1038/s41567-025-03090-6},
  note    = {arXiv:2501.04612}
}

@misc{wang2026sclsprocessor,
  title  = {A superconducting surface-code processor with lattice-surgery logical operations},
  author = {Yanzhe Wang and Fanhao Shen and Haipeng Xie and Aosai Zhang and Yu Gao and
            Chuanyu Zhang and Xuhao Zhu and Feitong Jin and Yiren Zou and Ning Wang and
            Zhengyi Cui and Zehang Bao and Zitian Zhu and Jiarun Zhong and Gongyu Liu and
            Jia-Nan Yang and Yihang Han and Yiyang He and Jiayuan Shen and Han Wang and
            Jiahua Huang and Xinrong Zhang and Sailang Zhou and Hang Dong and
            Jinfeng Deng and Yaozu Wu and Zixuan Song and Hekang Li and Zhen Wang and
            Chao Song and Qiujiang Guo and Pengfei Zhang and H. Wang and Ying Li},
  year   = {2026},
  note   = {arXiv:2606.06598}
}

@article{horsman2012lattice,
  title   = {Surface code quantum computing by lattice surgery},
  author  = {Dominic Horsman and Austin G. Fowler and Simon Devitt and Rodney {Van Meter}},
  journal = {New Journal of Physics},
  volume  = {14},
  number  = {12},
  pages   = {123011},
  year    = {2012},
  doi     = {10.1088/1367-2630/14/12/123011},
  note    = {arXiv:1111.4022}
}

@misc{fowler2018low,
  title  = {Low overhead quantum computation using lattice surgery},
  author = {Austin G. Fowler and Craig Gidney},
  note   = {arXiv:1808.06709},
  year   = {2018}
}

@article{litinski2019game,
  title   = {A Game of Surface Codes: Large-Scale Quantum Computing with Lattice Surgery},
  author  = {Daniel Litinski},
  journal = {Quantum},
  volume  = {3},
  pages   = {128},
  year    = {2019},
  doi     = {10.22331/q-2019-03-05-128},
  note    = {arXiv:1808.02892}
}

@article{litinski2019magic,
  title   = {Magic State Distillation: Not as Costly as You Think},
  author  = {Daniel Litinski},
  journal = {Quantum},
  volume  = {3},
  pages   = {205},
  year    = {2019},
  doi     = {10.22331/q-2019-12-02-205}
}

@article{bravyi2016trading,
  title   = {Trading Classical and Quantum Computational Resources},
  author  = {Sergey Bravyi and Graeme Smith and John A. Smolin},
  journal = {Physical Review X},
  volume  = {6},
  pages   = {021043},
  year    = {2016},
  doi     = {10.1103/PhysRevX.6.021043},
}

@article{beverland2022edpc,
  title   = {Surface Code Compilation via Edge-Disjoint Paths},
  author  = {Michael E. Beverland and Vadym Kliuchnikov and Eddie Schoute},
  journal = {PRX Quantum},
  volume  = {3},
  pages   = {020342},
  year    = {2022},
  doi     = {10.1103/PRXQuantum.3.020342},
  note    = {arXiv:2110.11493}
}

@article{gidney2024inplacey,
  title   = {Inplace Access to the Surface Code {Y} Basis},
  author  = {Craig Gidney},
  journal = {Quantum},
  volume  = {8},
  pages   = {1310},
  year    = {2024},
  doi     = {10.22331/q-2024-04-08-1310},
  note    = {arXiv:2302.07395}
}

@article{hamada2026routing,
  title={Efficient and high-performance routing of lattice-surgery paths on three-dimensional lattice},
  author={Hamada, Kou and Suzuki, Yasunari and Tokunaga, Yuuki},
  journal={Quantum},
  volume={10},
  pages={2061},
  year={2026},
  doi={10.22331/q-2026-04-13-2061},
  note={arXiv:2401.15829},
  publisher={Verein zur F{\"o}rderung des Open Access Publizierens in den Quantenwissenschaften}
}

@article{molavi2025dascot,
  title   = {Dependency-Aware Compilation for Surface Code Quantum Architectures},
  author  = {Abtin Molavi and Amanda Xu and Swamit Tannu and Aws Albarghouthi},
  journal = {Proceedings of the ACM on Programming Languages},
  volume  = {9},
  number  = {OOPSLA1},
  articleno = {82},
  numpages  = {28},
  month     = apr,
  year    = {2025},
  doi     = {10.1145/3720416},
  note    = {arXiv:2311.18042}
}

@misc{bolanos2025topologiq,
  title  = {{Topologiq}: Algorithmic Lattice Surgery},
  author = {Jose A. Bolanos and Austin Fowler},
  note   = {Software, \url{https://github.com/tqec/topologiq}},
  year   = {2025}
}

@misc{topols2026,
  title  = {{TopoLS}: Lattice Surgery Compilation via Topological Program Transformations},
  author = {Junyu Zhou and Yuhao Liu and Ethan Decker and Justin Kalloor and Mathias Weiden and Kean Chen and Costin Iancu and Gushu Li},
  note   = {arXiv:2601.23109},
  year   = {2026}
}

@misc{o3ls2026,
  title  = {{O3LS}: Optimizing Lattice Surgery via Automatic Layout Searching and Loose Scheduling},
  author = {Chenghong Zhu and Xian Wu and Jiahan Chen and Keming He and Junjie Wu and Xin Wang and Lingling Lao},
  note   = {arXiv:2604.15099},
  year   = {2026}
}

@misc{kan2025sparo,
  title  = {{SPARO}: Surface-code Pauli-based Architectural Resource Optimization for Fault-tolerant Quantum Computing},
  author = {Shuwen Kan and Zefan Du and Chenxu Liu and Meng Wang and Yufei Ding and
            Ang Li and Ying Mao and Samuel Stein},
  year   = {2025},
  note   = {arXiv:2504.21854}
}

@article{lao2018mapping,
  title   = {Mapping of lattice surgery-based quantum circuits on surface code architectures},
  author  = {Lingling Lao and Bas {van Wee} and Imran Ashraf and J. {van Someren} and Nader Khammassi and Koen Bertels and Carmen G. Almudever},
  journal = {Quantum Science and Technology},
  volume  = {4},
  number  = {1},
  pages   = {015005},
  year = {2019},
  doi     = {10.1088/2058-9565/aadd1a},
}

@misc{huggins2025flasq,
  title  = {The {FLuid} {Allocation} of {Surface} code {Qubits} ({FLASQ}) cost model for early fault-tolerant quantum algorithms},
  author = {William J. Huggins and Tanuj Khattar and Amanda Xu and Matthew Harrigan and Christopher Kang and Guang Hao Low and Austin Fowler and Nicholas C. Rubin and Ryan Babbush},
  note   = {arXiv:2511.08508},
  year   = {2025}
}

@misc{hao2025breakeven,
  title  = {Compilation Pipeline for Predicting Algorithmic Break-Even in an Early-Fault-Tolerant Surface Code Architecture},
  author = {Tianyi Hao and Joseph Sullivan and Sivaprasad Omanakuttan and Michael A. Perlin and Ruslan Shaydulin},
  note   = {arXiv:2511.20947},
  year   = {2025}
}

@misc{gidney2024cultivation,
  title  = {Magic State Cultivation: Growing {T} States as Cheap as {CNOT} Gates},
  author = {Craig Gidney and Noah Shutty and Cody Jones},
  note   = {arXiv:2409.17595},
  year   = {2024}
}

@misc{gidney2023hook,
  title  = {Cleaner Magic States with Hook Injection},
  author = {Craig Gidney},
  note   = {arXiv:2302.12292},
  year   = {2023}
}

@misc{cross2017openqasm,
  title  = {Open Quantum Assembly Language},
  author = {Andrew W. Cross and Lev S. Bishop and John A. Smolin and Jay M. Gambetta},
  note   = {arXiv:1707.03429},
  year   = {2017}
}

@article{gidney2021stim,
  title   = {Stim: a fast stabilizer circuit simulator},
  author  = {Craig Gidney},
  journal = {Quantum},
  volume  = {5},
  pages   = {497},
  year    = {2021},
  doi     = {10.22331/q-2021-07-06-497}
}

@article{tqec2026joss,
  title   = {{tqec}: A Python package for topological quantum error correction},
  author  = {Adrien Suau and Yiming Zhang and Purva Thakre and Yilun Zhao and
             Kabir Dubey and Jose A. Bolanos and Arabella Schelpe and Tianyi Hao and
             Philip Seitz and Gian Giacomo Guerreschi and
             {\'A}ngela Elisa {\'A}lvarez P{\'e}rez and Reinhard Stahn and
             Jerome Lenssen and Brendan Reid and Austin Fowler},
  journal = {Journal of Open Source Software},
  volume  = {11},
  number  = {120},
  pages   = {9142},
  year    = {2026},
  doi     = {10.21105/joss.09142}
}

@inproceedings{ding2020square,
  title={SQUARE: strategic quantum ancilla reuse for modular quantum programs via cost-effective uncomputation},
  author={Ding, Yongshan and Wu, Xin-Chuan and Holmes, Adam and Wiseth, Ash and Franklin, Diana and Martonosi, Margaret and Chong, Frederic T.},
  booktitle={2020 ACM/IEEE 47th Annual International Symposium on Computer Architecture (ISCA)},
  pages={570--583},
  year={2020},
  doi={10.1109/ISCA45697.2020.00054},
  organization={IEEE}
}

@article{xu2026bicycle,
  title={Distilling magic states in the bicycle architecture},
  author={Xu, Shifan and Liu, Kun and Rall, Patrick and He, Zhiyang and Ding, Yongshan},
  journal={arXiv preprint arXiv:2602.20546},
  year={2026}
}

@misc{puremagic2025,
  title  = {{PureMagic}: A Dynamic Scheduler for Lattice Surgery},
  author = {Hofmeyr, Steven and Weiden, Mathias and Kalloor, Justin and Kubiatowicz, John and Iancu, Costin},
  eprint = {2512.06484},
  year   = {2025}
}

@inproceedings{paykin2023pcoast,
  title={{PCOAST}: A {Pauli}-based quantum circuit optimization framework},
  author={Paykin, Jennifer and Schmitz, Albert T. and Ibrahim, Mohannad and Wu, Xin-Chuan and Matsuura, Anne Y.},
  booktitle={2023 IEEE International Conference on Quantum Computing and Engineering (QCE)},
  volume={1},
  pages={715--726},
  year={2023},
  organization={IEEE}
}

@inproceedings{schmitz2023pcoastext,
  title     = {Optimization at the Interface of Unitary and Non-unitary Quantum Operations in {PCOAST}},
  author    = {Schmitz, Albert T. and Ibrahim, Mohannad and Sawaya, Nicolas P. D. and Guerreschi, Gian Giacomo and Paykin, Jennifer and Wu, Xin-Chuan and Matsuura, Anne Y.},
  booktitle = {IEEE International Conference on Quantum Computing and Engineering (QCE)},
  eprint    = {2305.09843},
  pages     = {727--738},
  doi       = {10.1109/QCE57702.2023.00088},
  year      = {2023}
}

@article{peres2025pbcweights,
  title={Reducing depth and measurement weights in Pauli-based computation},
  author={Peres, Filipa C. R. and Galv{\~a}o, Ernesto F.},
  journal={Physical Review A},
  volume={112},
  number={6},
  pages={062604},
  year={2025},
  publisher={APS}
}

@article{sethi2026ppr,
  title   = {Optimizing Parallel Execution of Commuting {Pauli} Product Rotations},
  author  = {Sethi, Sayam and Nambisan, Devika and Baker, Jonathan Mark},
  journal = {arXiv preprint arXiv:2605.23738},
  year    = {2026}
}

@article{bharadwaj2026qomet,
  title   = {{C-Phase}-Aware Compilation for Efficient Fault-Tolerant Quantum Execution},
  author  = {Bharadwaj, Dhanvi and Dangwal, Siddharth and Hou, Yuewen and Ravi, Gokul Subramanian},
  journal = {arXiv preprint arXiv:2605.14042},
  year    = {2026}
}

@article{coecke2011interacting,
  title={Interacting quantum observables: categorical algebra and diagrammatics},
  author={Coecke, Bob and Duncan, Ross},
  journal={New Journal of Physics},
  volume={13},
  number={4},
  pages={043016},
  year={2011}
}

@article{vandewetering2020zxcalculus,
  title   = {{ZX}-calculus for the working quantum computer scientist},
  author  = {van de Wetering, John},
  journal = {arXiv preprint arXiv:2012.13966},
  year    = {2020}
}

@inproceedings{wang2026transpiler,
  title={Transpiler-Architecture Co-Design to Curb Clifford Costs in Fault-Tolerant Quantum Computing},
  author={Wang, Meng and Liu, Chenxu and Stein, Samuel and Ding, Yufei and Das, Poulami and Nair, Prashant J. and Li, Ang},
  booktitle={2026 ACM/IEEE 53rd Annual International Symposium on Computer Architecture (ISCA)},
  pages={889--904},
  year={2026},
  organization={IEEE}
}

@article{wang2025tableau,
  title   = {Tableau-Based Framework for Efficient Logical Quantum Compilation},
  author  = {Wang, Meng and Liu, Chenxu and Garner, Sean and Stein, Samuel and Ding, Yufei and Nair, Prashant J. and Li, Ang},
  journal = {arXiv preprint arXiv:2509.02721},
  year    = {2025}
}

@inproceedings{kissinger2020pyzx,
  title     = {{PyZX}: Large Scale Automated Diagrammatic Reasoning},
  author    = {Kissinger, Aleks and van de Wetering, John},
  booktitle = {Proceedings of the 16th International Conference on Quantum Physics and Logic (QPL)},
  series    = {EPTCS},
  volume    = {318},
  pages     = {229--241},
  year      = {2020},
  note      = {arXiv:1904.04735}
}

@phdthesis{gottesman1997stabilizer,
  author = {Gottesman, Daniel},
  title  = {Stabilizer Codes and Quantum Error Correction},
  school = {California Institute of Technology},
  year   = {1997},
  note   = {arXiv:quant-ph/9705052}
}

@inproceedings{cuthill1969reducing,
  title     = {Reducing the bandwidth of sparse symmetric matrices},
  author    = {Cuthill, Elizabeth and McKee, James},
  booktitle = {Proceedings of the 24th ACM National Conference},
  pages     = {157--172},
  year      = {1969}
}

@article{hilbert1891curve,
  title   = {{\"U}ber die stetige {Abbildung} einer {Linie} auf ein {Fl{\"a}chenst{\"u}ck}},
  author  = {Hilbert, David},
  journal = {Mathematische Annalen},
  volume  = {38},
  pages   = {459--460},
  year    = {1891}
}

@article{dijkstra1959note,
  title   = {A note on two problems in connexion with graphs},
  author  = {Dijkstra, Edsger W.},
  journal = {Numerische Mathematik},
  volume  = {1},
  pages   = {269--271},
  year    = {1959}
}

@misc{lightstim2026autodem,
  title  = {{LightStim}: A Framework for {QEC} Protocol Evaluation and Prototyping with Automated {DEM} Construction},
  author = {Fang, Xiang and Wang, Ming and Wu, Yue and Prabhu, Sharanya and Tullsen, Dean and Miniskar, Narasinga Rao and Mueller, Frank and Humble, Travis and Ding, Yufei},
  eprint = {2604.21472},
  year   = {2026},
}

@misc{kishony2026surface,
  title  = {Surface code off-the-hook: diagonal syndrome-extraction scheduling},
  author = {Kishony, Gilad and Fowler, Austin},
  note   = {arXiv:2602.09099},
  year   = {2026}
}

@article{chamberland2022twistfree,
  title   = {Universal quantum computing with twist-free and temporally encoded lattice surgery},
  author  = {Christopher Chamberland and Earl T. Campbell},
  journal = {PRX Quantum},
  volume  = {3},
  pages   = {010331},
  year    = {2022},
  doi     = {10.1103/PRXQuantum.3.010331},
  note    = {arXiv:2109.02746}
}

@article{geher2024hadamard,
  title   = {Error-corrected {Hadamard} gate simulated at the circuit level},
  author  = {Gy{\"o}rgy P. Geh{\'e}r and Campbell McLauchlan and Earl T. Campbell and Alexandra E. Moylett and Ophelia Crawford},
  journal = {Quantum},
  volume  = {8},
  pages   = {1394},
  year    = {2024},
  doi     = {10.22331/q-2024-07-02-1394},
  note    = {arXiv:2312.11605}
}

@article{li2023qasmbench,
  title={{QASMBench}: A low-level quantum benchmark suite for {NISQ} evaluation and simulation},
  author={Li, Ang and Stein, Samuel and Krishnamoorthy, Sriram and Ang, James},
  journal={ACM Transactions on Quantum Computing},
  volume={4},
  number={2},
  pages={1--26},
  year={2023},
  publisher={ACM New York, NY}
}

@article{higgott2025sparseblossom,
  author  = {Higgott, Oscar and Gidney, Craig},
  title   = {Sparse Blossom: correcting a million errors per core second with minimum-weight matching},
  journal = {Quantum},
  volume  = {9},
  pages   = {1600},
  year    = {2025},
  doi     = {10.22331/q-2025-01-20-1600}
}

@misc{beverland2022req,
  title  = {Assessing requirements to scale to practical quantum advantage},
  author = {Michael E. Beverland and Prakash Murali and Matthias Troyer and
            Krysta M. Svore and Torsten Hoefler and Vadym Kliuchnikov and
            Guang Hao Low and Mathias Soeken and Aarthi Sundaram and
            Alexander Vaschillo},
  year   = {2022},
  note   = {arXiv:2211.07629}
}

@article{erickson1987sendsplit,
  title   = {Send-and-Split Method for Minimum-Concave-Cost Network Flows},
  author  = {Erickson, Ranel E. and Monma, Clyde L. and Veinott, Jr., Arthur F.},
  journal = {Mathematics of Operations Research},
  volume  = {12},
  number  = {4},
  pages   = {634--664},
  year    = {1987},
  doi     = {10.1287/moor.12.4.634}
}

@misc{robertson2025resource,
  title  = {A Resource Allocating Compiler for Lattice Surgery},
  author = {Alan Robertson and Haowen Gao and Yuval R. Sanders},
  year   = {2025},
  note   = {arXiv:2506.04620},
}

@article{pflieger2026harvest,
  title={Harvest: Resource-Aware Quantum Compilation for Magic State Protocols},
  author={Pflieger, Jannik and {\'S}wierkowska, Aleksandra and Giortamis, Emmanouil and Bhatotia, Pramod},
  journal={arXiv preprint arXiv:2608.03315},
  year={2026}
}

\end{document}